\documentclass[10pt]{article}
\PassOptionsToPackage{dvipsnames,table}{xcolor}
\PassOptionsToPackage{numbers,sort&compress}{natbib}
\usepackage[wide,nowritingstyle]{preprint}
\usepackage[utf8]{inputenc}
\usepackage{microtype}
\usepackage{pifont,xspace,amsthm,fontawesome5}
\usepackage[most]{tcolorbox}
\usepackage{wrapfig,enumitem,subcaption,multirow,makecell}
\usepackage{algorithm,algpseudocode,hhline,tablefootnote}
\usepackage{adjustbox}
\usepackage[bottom]{footmisc}
\usepackage{hyperref}
\usepackage{pricoder-layout}

\definecolor{TsinghuaPurple}{RGB}{102,8,116}
\newcommand{\lrevision}[1]{#1}
\newenvironment{prevision}{\par}{\par}
\newcommand{\crevision}{}

\newcommand{\seclrevision}[1]{#1}
\newenvironment{secprevision}{\par}{\par}
\newcommand{\seccrevision}{}

\algrenewcommand\algorithmicrequire{\textbf{Input:}}
\algrenewcommand\algorithmicensure{\textbf{Output:}}

\def\name{\textit{\textsc{PriCoder}}\xspace}

\def\ndonnx{{\fontfamily{cmtt}\selectfont{NdonnxEval}}\xspace}
\def\numba{{\fontfamily{cmtt}\selectfont{NumbaEval}}\xspace}
\def\humaneval{{\fontfamily{cmtt}\selectfont{HumanEval}}\xspace}
\def\mbpp{{\fontfamily{cmtt}\selectfont{MBPP}}\xspace}

\def\qwen{{\fontfamily{pcr}\selectfont{Qwen-7B}}\xspace}
\def\deepseek{{\fontfamily{pcr}\selectfont{DeepSeek-6.7B}}\xspace}
\def\llama{{\fontfamily{pcr}\selectfont{LLaMA-8B}}\xspace}

\newcommand{\bi}[1]{\textbf{\textit{#1}}}
\newcommand{\pass}[1]{\textit{pass@#1}}
\newcommand{\exec}[1]{\textit{exec@#1}}
\newcommand{\recall}[1]{\textit{recall@#1}}

\newcommand{\baselinename}[1]{\textit{#1}\xspace}
\newcommand{\llmname}[1]{{\fontfamily{pcr}\selectfont {#1}}\xspace}
\newcommand{\dataname}[1]{{\fontfamily{cmtt}\selectfont {#1}}\xspace}

\definecolor{namerow}{RGB}{246,249,255}

\definecolor{vanillarow}{RGB}{252,252,252} 
\definecolor{ragrow}{RGB}{246,252,255}     
\definecolor{newnamerow}{RGB}{246,253,246} 

\logos{%
  \mbox{%
    \raisebox{-0.5\height}{%
      \includegraphics[height=0.30in,
        trim=465bp 80bp 490bp 75bp,clip]{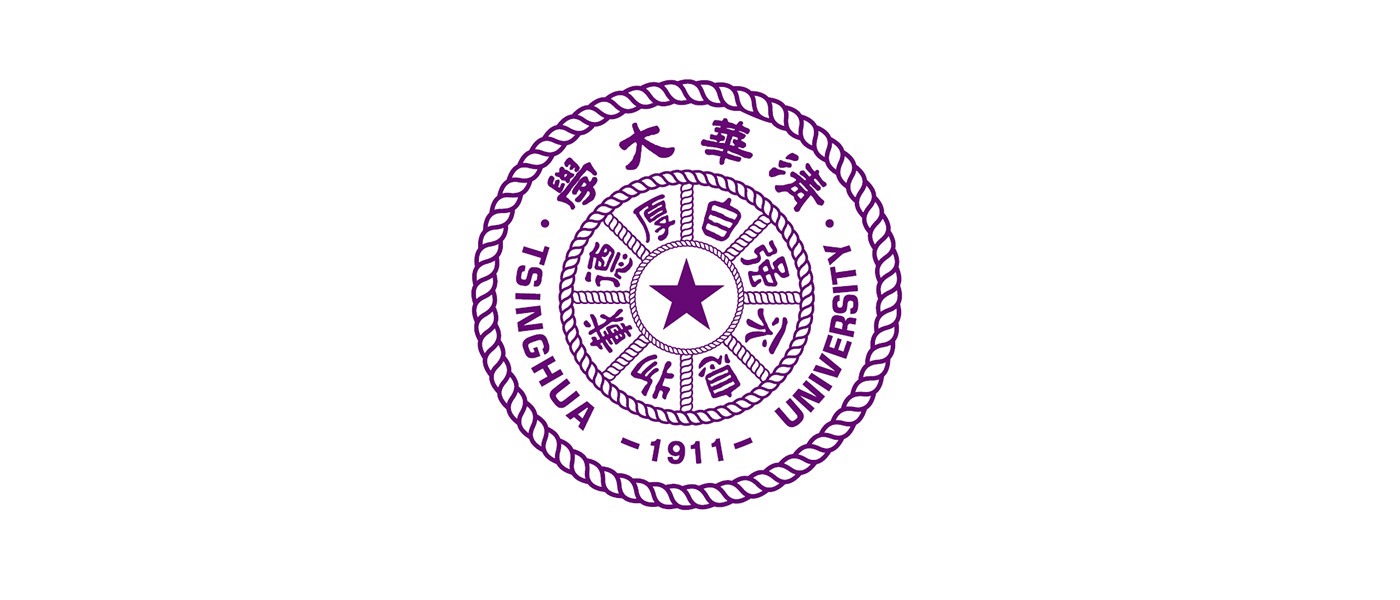}}%
    \hspace{7pt}%
    \raisebox{-0.5\height}{%
      \includegraphics[height=0.32in,
        trim=0bp 0bp 0bp 120bp,clip]{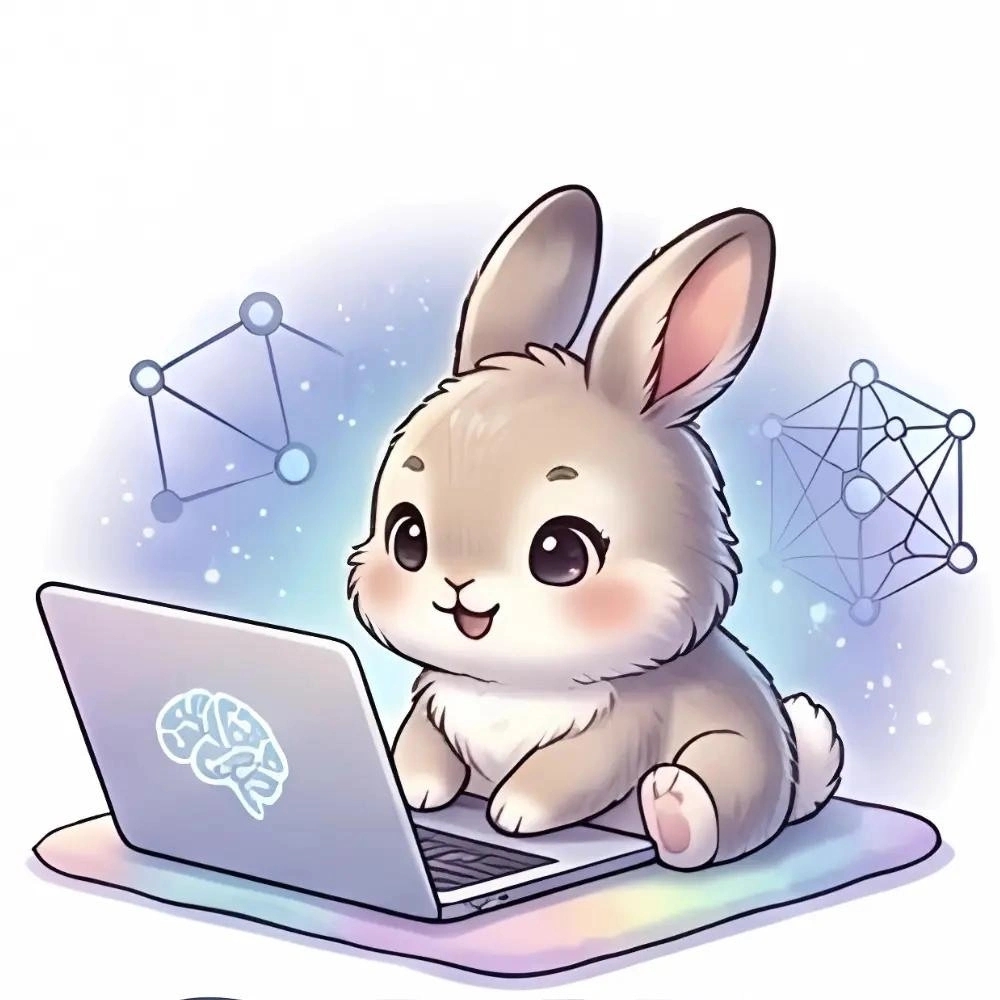}}%
    \hspace{7pt}%
    \raisebox{-0.5\height}{%
      \fontfamily{Montserrat-LF}\fontseries{medium}\fontshape{n}%
      \fontsize{10.5}{12}\selectfont\color{accent}%
      \textls[10]{\groupname}}%
  }%
}

\title{To See is Not to Master: Teaching LLMs to Use\\Private Libraries for Code Generation}
\author[1,2,*]{\href{https://orcid.org/0009-0000-1138-4503}{\textcolor{black}{Yitong Zhang}}}
\author[3,*]{\href{https://orcid.org/0009-0008-0431-0125}{\textcolor{black}{Chengze Li}}}
\author[4,*]{\href{https://orcid.org/0009-0009-7149-1998}{\textcolor{black}{Ruize Chen}}}
\author[2]{\href{https://orcid.org/0000-0003-4801-9985}{\textcolor{black}{Guowei Yang}}}
\author[5]{\href{https://orcid.org/0009-0005-8133-2502}{\textcolor{black}{Xiaoran Jia}}}
\author[6]{\href{https://orcid.org/0009-0008-4799-7270}{\textcolor{black}{Yijie Ren}}}
\author[1,\dagger]{\href{https://orcid.org/0000-0002-5579-8852}{\textcolor{black}{Jia Li}}}

\affiliation[1]{College of AI, Tsinghua University, Beijing, China}
\affiliation[2]{Fitten Tech Co., Ltd., Beijing, China}
\affiliation[3]{School of Computer Science, Nanjing University, Nanjing, China}
\affiliation[4]{Software Institute, Nanjing University, Nanjing, China}
\affiliation[5]{School of Computer Science and Technology, Beijing Institute of Technology, Beijing, China}
\affiliation[6]{School of Computer Science and Engineering, Beihang University, Beijing, China}

\metadata[Yitong Zhang]{\email{zhangyt26@mails.tsinghua.edu.cn}}
\metadata[Jia Li]{\email{jia\_li@mail.tsinghua.edu.cn}}

\newcommand{\paperauthornotes}{%
  \begingroup
  \renewcommand{\thefootnote}{\fnsymbol{footnote}}%
  \footnotetext[1]{Equal contribution. \ding{182} Yitong Zhang and Chengze Li jointly conceived the idea. \ding{183} Yitong Zhang took the lead in writing the paper. \ding{184} Chengze Li implemented the proposed method. \ding{185} Ruize Chen contributed to baseline reproduction.}%
  \footnotetext[2]{Corresponding author.}%
  \endgroup
  \setcounter{footnote}{0}%
}

\runninghead{\groupname \hfill PriCoder}
\hypersetup{
  pdftitle={To See is Not to Master: Teaching LLMs to Use Private Libraries for Code Generation},
  pdfauthor={Yitong Zhang; Chengze Li; Ruize Chen; Guowei Yang; Xiaoran Jia; Yijie Ren; Jia Li},
  pdfkeywords={Private-Library-Oriented Code Generation, Large Language Models},
  pdfsubject={Software libraries and repositories; Neural networks; Automatic programming; Natural language processing}
}

\begin{document}
\maketitle
\paperauthornotes
\begin{abstract}
Large Language Models (LLMs) have shown strong potential for code generation, yet they remain limited in private-library-oriented code generation, where models are expected to generate code that correctly invokes APIs from private libraries. Existing approaches primarily rely on retrieving private-library API documentation and injecting the retrieved knowledge into the context at inference time. However, our study shows that this strategy is insufficient: \emph{even when provided with accurate required API knowledge, LLMs still struggle to invoke private-library APIs effectively}.

To address this limitation, we propose \name, an approach that improves LLMs' ability to \emph{invoke} private-library APIs through automatically synthesized training data. \lrevision{Specifically, \name uses a synthesis graph to organize private-library data synthesis and applies two complementary operations: \ding{182} \emph{Progressive Graph Evolution}, which improves data diversity by progressively synthesizing more complex and diverse training samples from basic ones, and \ding{183} \emph{Multidimensional Graph Pruning}, which improves data quality through an automatic filtering pipeline.}
To support rigorous evaluation, we construct two new benchmarks based on recently released libraries that are unfamiliar to the evaluated models. \lrevision{Across the two private-library benchmarks and all evaluated models, \name improves over \baselinename{Vanilla} by 13.53--31.66 percentage points in \pass{k}. On \humaneval and \mbpp, \name generally achieves higher performance than RAG-based baselines.}
\end{abstract}

{\small\textbf{Keywords:} Private-Library-Oriented Code Generation, Large Language Models\par}

\section{Introduction}
\label{sec:intro}

Recently, Large Language Models (LLMs) have shown strong potential for code generation~\cite{li2025structured, jiang2025aixcoder, cai2025ai, li2025beyond, zhang2026grammar}. However, many real-world development scenarios rely heavily on internal private libraries, which are rarely included in public training corpora~\cite{apifinder, exploracoder}. This gives rise to an important yet challenging task, namely \textit{Private-Library-Oriented Code Generation}, which aims to automatically generate code that invokes private-library APIs to satisfy specific coding requirements. Since LLMs typically lack prior knowledge of such libraries, they often struggle to use private libraries effectively, which largely restricts the practical impact of LLMs in real-world software development.

To mitigate this deficiency, a growing body of work has explored private-library-oriented code generation through Retrieval-Augmented Generation (RAG)~\cite{apifinder, epigen, docprompting, exploracoder}. These approaches typically retrieve relevant API specifications from documentation and inject them into the context. They implicitly assume that, once the model is presented with the required knowledge, it can reliably invoke the required APIs to satisfy the coding requirement. However, we notice that a fundamental question remains underexplored: \bi{given the required API knowledge, can LLMs actually utilize it effectively?} To answer this question, we conduct an empirical study in Section~\ref{sec:motivation}. Our results show that even when provided with detailed specifications of all required APIs, LLMs still struggle to invoke these APIs effectively to satisfy the coding requirements. 
For example, even when equipped with the complete set of required private APIs, the \pass{1} of LLaMA3.1-8B-Instruct~\cite{llama} improves only from 8.13\% to 13.10\%, which remains far from practical usability.
\lrevision{Larger models achieve higher absolute performance, but gains from oracle API specifications remain limited: LLaMA3.1-70B-Instruct~\cite{llama} improves only from 35.88\% to 44.22\% in \pass{1} on \numba.}
These findings suggest that the key bottleneck is not merely whether the model can \emph{see} the right API knowledge before generation, but whether it can \emph{invoke} private-library APIs correctly during generation. 
Therefore, in this paper, different from most prior work~\cite{epigen, docprompting, capir}, we focus on how to improve LLMs’ ability to effectively \emph{invoke} private-library APIs.

One fundamental reason why LLMs perform poorly on private libraries is that such libraries are typically absent from the training corpora~\cite{apifinder}. This naturally suggests training LLMs on data related to the target private library.
A natural next question, then, is how to obtain training data about private libraries at scale. Due to the closed-source nature of private libraries, real-world code that invokes them is often scarce, while manually curating such data is labor-intensive and impractical~\cite{epigen, majumdar2025genetic}.
We therefore turn to LLM-driven data synthesis, where LLMs automatically generate training samples that are subsequently used to fine-tune themselves~\cite{luo2023wizardcoder, liu2026self, wei2023magicoder, wu2025ucoder}.
However, direct synthesis is far from trivial. As shown in Section~\ref{sec:rq3}, fine-tuning models on directly synthesized data yields a marginal improvement of less than 10 percentage points in \pass{1}. This limited performance gain is primarily because models initially lack sufficient prior knowledge of the target private library, causing the directly synthesized data to suffer from two major challenges.
\ding{182} \bi{Low Data Diversity.} LLMs tend to generate overly basic requirements that can be solved with only a small number of private APIs. In contrast, real-world development scenarios often require the coordinated invocation of multiple private APIs to satisfy diverse and complex coding requirements. This creates a substantial gap between synthesized data and practical development needs. \ding{183} \bi{Poor Data Quality.} LLMs are prone to producing flawed training samples, such as those containing syntax errors, non-existent APIs, or semantically incorrect API invocations. Directly training on such noisy data can be counterproductive and may even degrade the model’s overall capabilities.

To address these challenges, we propose \name, an approach that teaches LLMs to \emph{invoke} private-library APIs. 
The main idea of \name is to let LLMs automatically synthesize training samples tailored to the target private library and then learn private-library API invocation through training.
\lrevision{Specifically, \name uses a graph to organize the synthesis process and applies two operators to generate and filter samples from private API specifications:}
\ding{182} First, \bi{Progressive Graph Evolution} improves data diversity by progressively synthesizing new, diverse sample nodes based on existing basic ones. \lrevision{Through initial and iterative evolution, this operator expands the range of API compositions represented in the synthesized data.}
\ding{183} Second, \bi{Multidimensional Graph Pruning} ensures data quality by removing low-quality sample nodes through a multidimensional verification pipeline: we (i) eliminate samples with syntactic errors, (ii) validate executability via automatically generated tests, and (iii) further assess overall functionality automatically. 
\lrevision{Together, these two operators construct a diverse and reliable synthetic dataset that enables LLMs to learn private-library API invocation without human intervention.}

Evaluating private-library-oriented code generation requires libraries that are unseen to the model. However, many existing benchmarks are built on libraries that have been publicly released for a long time, making them unreliable for assessing the private-library-oriented generation capability of today's popular LLMs~\cite{apifinder, exploracoder}.
To obtain a more faithful evaluation, we select two recently released libraries, \texttt{ndonnx}~\cite{ndonnx} and \texttt{numba-cuda}~\cite{numba}, both of which were introduced in 2024 and continued to largely evolve throughout 2025. Based on them, we carefully construct two new benchmarks, namely \ndonnx and \numba, which contain 169 and 187 instances, respectively.
\lrevision{We evaluate \name on multiple LLMs and compare it with approaches that provide API specifications, source code, or retrieved usage examples.}
\lrevision{Across the two private-library benchmarks and all evaluated models, \name improves over \baselinename{Vanilla} by 13.53--31.66 percentage points in \pass{k}.}
\lrevision{On \humaneval and \mbpp, \name generally achieves higher performance than RAG-based baselines.} Ablation studies further confirm that both Progressive Graph Evolution and Multidimensional Graph Pruning are essential to the overall effectiveness of our approach.

In summary, this paper makes the following contributions:
\begin{itemize}[leftmargin=12pt]
\item We empirically show that LLMs still struggle to correctly invoke private-library APIs for code generation, even when the required API knowledge is explicitly provided in the context.
\item We propose \name, a data-synthesis and training framework that automatically constructs library-specific training data and improves LLMs' ability to invoke private-library APIs.
\item We construct and release two new benchmarks for private-library-oriented code generation. Extensive experiments on multiple LLMs demonstrate that \lrevision{\name improves private-library-oriented code generation and generally outperforms RAG-based baselines on \humaneval and \mbpp.}
\end{itemize}

\section{Background and Related Work}
\label{sec:related}

\subsection{Large Language Models for Code Generation}
\label{sec:related-llm}

Large language models (LLMs) have recently become a powerful foundation for code generation, substantially advancing the automation of software development~\cite{qian2024chatdev, li2026papers, zhang2026lookahead, li2024acecoder, yang2025difftester, chen2026coact}.
Recent model families, such as Qwen~\cite{yang2025qwen3, qwencoder}, DeepSeek~\cite{liu2024deepseek, deepseekcoder}, LLaMA~\cite{llama, roziere2023code}, and GPT~\cite{singh2025openai, hurst2024gpt}, have achieved strong performance on a wide range of code generation benchmarks.
These results suggest that modern LLMs can learn common programming patterns, language syntax, and usage conventions of widely adopted libraries.
However, their behavior remains strongly shaped by the knowledge observed during training~\cite{wang2025codesync}.
After training, a model may struggle to reliably solve tasks that depend on non-public, newly released, or rapidly evolving library knowledge~\cite{yuan2026se, jiang2026koco, nashid2024contextual}.

\subsection{Private-Library-Oriented Code Generation}
\label{sec:related-pri}

Private-Library-Oriented Code Generation aims to generate code that satisfies a coding requirement by invoking APIs from a target private library~\cite{apifinder}.
This setting is more challenging than conventional library-oriented code generation~\cite{zan2024diffcoder, gu2025effectiveness, liu2025think, liu2023codegen4libs}, because the target libraries are often invisible during pretraining and their usage patterns are rarely documented through public code examples.
In many industrial environments, such libraries encode organization-specific abstractions, data structures, and development conventions; as a result, a model must not only understand a task requirement, but also determine how to realize it through APIs whose semantics and invocation patterns are largely unfamiliar.

Existing studies mainly address this problem through Retrieval-Augmented Generation (RAG)~\cite{lewis2020retrieval}.
APIFinder~\cite{apifinder} and DocPrompting~\cite{docprompting} retrieve relevant API documentation and inject the retrieved specifications into the prompt.
EpiGen~\cite{epigen} and CAPIR~\cite{capir} further improve retrieval by decomposing a coding requirement into subtasks before retrieving and reranking API information.
ExploraCoder~\cite{exploracoder} incorporates execution feedback to reduce errors caused by ambiguous or incomplete API documentation.
Overall, most existing approaches focus on improving the accuracy of retrieved API knowledge, enabling the model to \bi{see} more relevant information at inference time. However, they pay relatively limited attention to whether the model can actually \bi{invoke} private-library APIs correctly and compose them to satisfy the intended functionality.
\section{Motivation}
\label{sec:motivation}

\begin{table}[!htbp]
\centering
\small
\renewcommand{\arraystretch}{0.87}
\setlength{\tabcolsep}{12pt}
\caption{\textit{\pass{k}} (\%) on \numba and \ndonnx.}
\label{tab:motivation}

\begin{adjustbox}{max width=\linewidth}
\begin{tabular}{c|ccc|ccc}
\toprule
\multirow{2}{*}{\textbf{Method}} &
\multicolumn{3}{c|}{\textbf{\numba}} &
\multicolumn{3}{c}{\textbf{\ndonnx}} \\
\cmidrule(lr){2-4}\cmidrule(lr){5-7}
& \textbf{\pass{1}} & \textbf{\pass{3}} & \textbf{\pass{5}}
& \textbf{\pass{1}} & \textbf{\pass{3}} & \textbf{\pass{5}} \\
\midrule

\multicolumn{7}{c}{\textbf{\llmname{LLaMA3.1-8B-Instruct}}} \\
\arrayrulecolor{gray}\midrule\arrayrulecolor{black}
\baselinename{Vanilla} & 8.13 & 16.97 & 21.73 & 15.86 & 23.50 & 26.55 \\
\baselinename{Oracle}  & 13.10 & 26.73 & 33.66 & 30.36 & 42.79 & 47.45 \\
\cellcolor[HTML]{E9F7EF}\textbf{Gain} (\(\uparrow\))
& \cellcolor[HTML]{E9F7EF}\textbf{+4.97} & \cellcolor[HTML]{E9F7EF}\textbf{+9.76} & \cellcolor[HTML]{E9F7EF}\textbf{+11.93}
& \cellcolor[HTML]{E9F7EF}\textbf{+14.50} & \cellcolor[HTML]{E9F7EF}\textbf{+19.29} & \cellcolor[HTML]{E9F7EF}\textbf{+20.90} \\
\midrule

\multicolumn{7}{c}{\textbf{\llmname{DeepSeek-Coder-6.7B-Instruct}}} \\
\arrayrulecolor{gray}\midrule\arrayrulecolor{black}
\baselinename{Vanilla} & 10.59 & 21.26 & 27.82 & 20.36 & 29.16 & 33.26 \\
\baselinename{Oracle}  & 17.97 & 34.23 & 42.45 & 38.05 & 50.82 & 55.86 \\
\cellcolor[HTML]{E9F7EF}\textbf{Gain} (\(\uparrow\))
& \cellcolor[HTML]{E9F7EF}\textbf{+7.38} & \cellcolor[HTML]{E9F7EF}\textbf{+12.97} & \cellcolor[HTML]{E9F7EF}\textbf{+14.63}
& \cellcolor[HTML]{E9F7EF}\textbf{+17.69} & \cellcolor[HTML]{E9F7EF}\textbf{+21.66} & \cellcolor[HTML]{E9F7EF}\textbf{+22.60} \\
\midrule

\multicolumn{7}{c}{\textbf{\llmname{LLaMA3.1-70B-Instruct}}} \\
\arrayrulecolor{gray}\midrule\arrayrulecolor{black}
\baselinename{Vanilla} & 35.88 & 52.41 & 59.28 & 27.99 & 36.07 & 39.34 \\
\baselinename{Oracle}  & 44.22 & 57.16 & 62.08 & 48.58 & 58.83 & 60.29 \\
\cellcolor[HTML]{E9F7EF}\textbf{Gain} (\(\uparrow\))
& \cellcolor[HTML]{E9F7EF}\textbf{+8.34}
& \cellcolor[HTML]{E9F7EF}\textbf{+4.75} 
& \cellcolor[HTML]{E9F7EF}\textbf{+2.80}
& \cellcolor[HTML]{E9F7EF}\textbf{+20.59}
& \cellcolor[HTML]{E9F7EF}\textbf{+22.76} 
& \cellcolor[HTML]{E9F7EF}\textbf{+20.95} \\
\midrule

\multicolumn{7}{c}{\textbf{\llmname{DeepSeek-V3-671B}}} \\
\arrayrulecolor{gray}\midrule\arrayrulecolor{black}
\baselinename{Vanilla} & 29.95 & 34.22 & 37.69 & 27.22 & 30.77 & 32.95 \\
\baselinename{Oracle}  & 39.04 & 42.25 & 45.45 & 37.87 & 42.60 & 46.15 \\
\cellcolor[HTML]{E9F7EF}\textbf{Gain} (\(\uparrow\))
& \cellcolor[HTML]{E9F7EF}\textbf{+9.09} 
& \cellcolor[HTML]{E9F7EF}\textbf{+8.03} 
& \cellcolor[HTML]{E9F7EF}\textbf{+7.76}
& \cellcolor[HTML]{E9F7EF}\textbf{+10.65}
& \cellcolor[HTML]{E9F7EF}\textbf{+11.83} 
& \cellcolor[HTML]{E9F7EF}\textbf{+13.20} \\

\bottomrule
\end{tabular}
\end{adjustbox}

\end{table}

Most prior work on private-library-oriented code generation is based on RAG. They implicitly assume that once the model is presented with the required knowledge, it can reliably invoke required APIs to satisfy the coding requirement. In this section, we challenge this assumption and show that \emph{seeing} the sufficient required API knowledge does not necessarily translate into \emph{invoking} it correctly.

To disentangle the impact of retrieval quality, we design an \baselinename{oracle} setting that eliminates potential retrieval errors.
Specifically, for each problem, we construct an \baselinename{oracle} prompt that includes the full specifications of all APIs required to solve the task, including their signatures and clear functional descriptions. \lrevision{This \baselinename{oracle} setting represents an idealized upper bound for documentation-based RAG approaches that use API specifications as inference-time context.}
We compare it with a \baselinename{vanilla} setting, which solves the same requirement but does not provide the model with any API documentation.
To ensure that the motivation study evaluates private-library usage rather than memorized library knowledge, we select models whose pre-training data is unlikely to include the target library versions used in our benchmarks.
Experiments are conducted on two benchmarks, \numba and \ndonnx (described in Section~\ref{sec:benchmark-construct}), using DeepSeek-Coder-6.7B-Instruct~\cite{deepseekcoder}, DeepSeek-V3-671B~\cite{liu2024deepseek}, LLaMA3.1-8B-Instruct~\cite{llama}, and LLaMA3.1-70B-Instruct~\cite{llama}.

\begin{wrapfigure}[32]{r}{0.47\textwidth}
    \centering
    \captionsetup{font=footnotesize}
    \includegraphics[width=\linewidth]{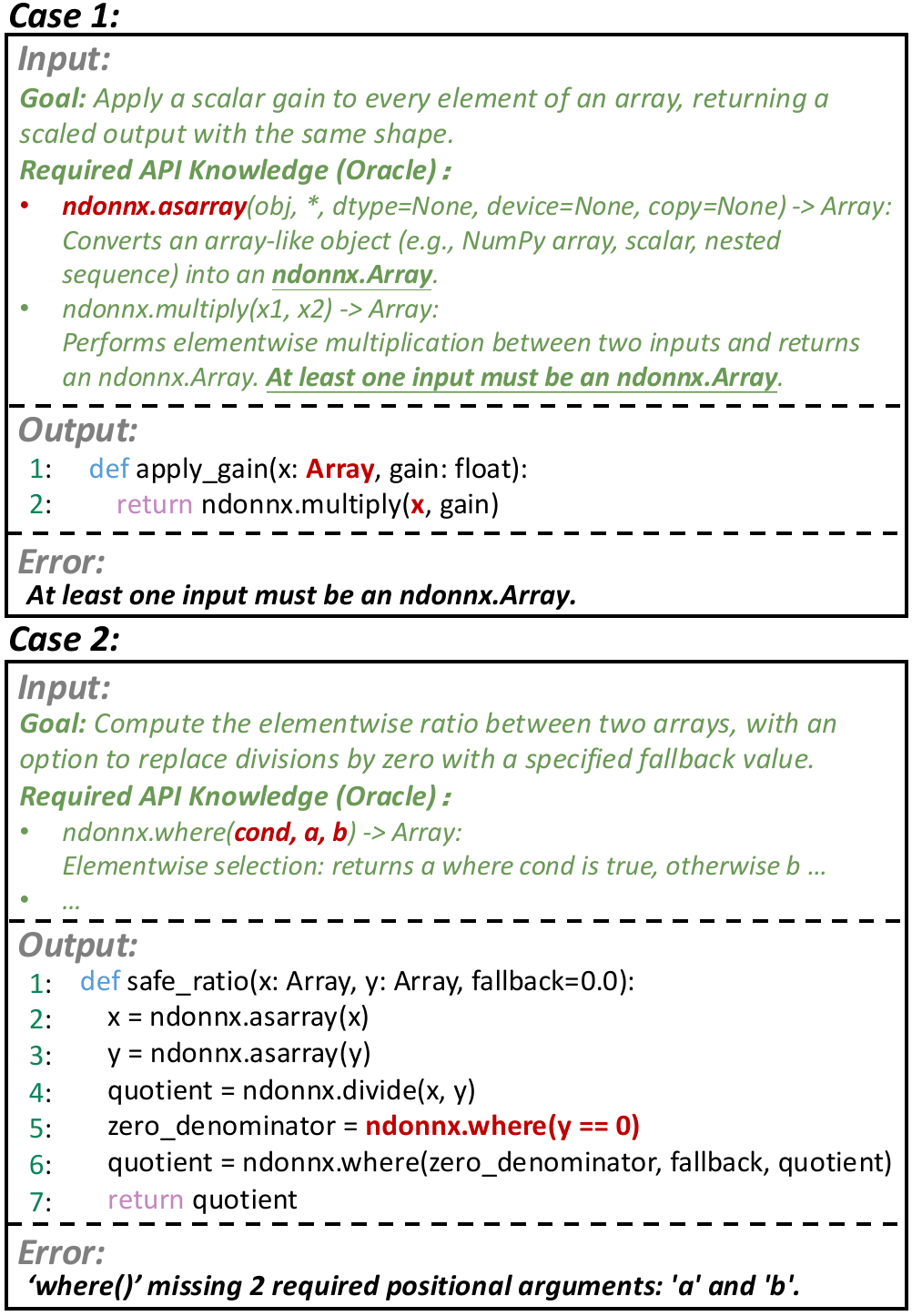}

    \caption{Two cases on \ndonnx with DeepSeek-Coder-6.7B-Instruct. Despite having access to oracle required API knowledge, the model fails to invoke these APIs effectively.}
    \label{fig:motivation}

\end{wrapfigure}

Table~\ref{tab:motivation} reports \pass{k} on \numba and \ndonnx. Although providing perfect API specifications improves performance, the absolute gains remain far from sufficient for practical deployment. For instance, on \numba, \textit{oracle} prompting only increases \pass{1} from 10.59\% to 17.97\% for {DeepSeek-Coder-6.7B-Instruct}, and from 8.13\% to 13.10\% for {LLaMA3.1-8B-Instruct}, leaving most instances unsolved. \lrevision{Larger models achieve higher absolute \pass{1} on \numba, but the gap between \baselinename{Vanilla} and \baselinename{Oracle} remains: for {LLaMA3.1-70B-Instruct}, \pass{1} increases from 35.88\% to 44.22\%, and for {DeepSeek-V3-671B}, it increases from 29.95\% to 39.04\%.}
\lrevision{Overall, these results suggest that the gains achievable through documentation-only prompting remain limited in the evaluated settings.}~\footnote{The non-zero \pass{k} under the \baselinename{vanilla} setting does not necessarily indicate that the model has mastered the target private libraries. Our case study shows that some instances can be solved without invoking private-library APIs, by implementing equivalent functionality with more standalone code.}
 
We further analyze the failed cases and find that errors are largely attributable to \emph{ineffective invocation} of the provided oracle APIs, rather than missing any required knowledge in the context. In many instances, the model either does not invoke a necessary API even when it is explicitly required, or invokes it in an incorrect manner.
We highlight two representative cases in Figure~\ref{fig:motivation}. 
\ding{192} In the first case, the task requires converting an array into an \colorbox{gray!15}{\texttt{ndonnx.Array}} via \colorbox{gray!15}{\texttt{ndonnx.asarray}} before applying \colorbox{gray!15}{\texttt{ndonnx.multiply}}. Despite being provided with both API specifications, the model directly calls \colorbox{gray!15}{\texttt{ndonnx.multiply}} on the raw input and omits the required conversion, which triggers a type error at runtime. 
\ding{193} In the second case, the task requires safely handling division-by-zero by using \colorbox{gray!15}{\texttt{ndonnx.where}} to select between the computed quotient and a fallback value. Although the model attempts to use \colorbox{gray!15}{\texttt{ndonnx.where}}, it misinterprets the API signature and calls \colorbox{gray!15}{\texttt{ndonnx.where}} with only one argument, leading to an argument error.

\begin{boxK}
\small \faIcon{pencil-alt} \textbf{Motivation 1:}
Even with oracle API knowledge, LLMs frequently fail to invoke APIs actively and correctly. 
The key bottleneck is enabling LLMs to \emph{invoke} private-library APIs effectively, rather than merely \emph{seeing} required API knowledge in context.
\end{boxK}

Motivated by the above findings, we argue that LLMs should be trained to learn how to invoke private-library APIs. However, training data for private libraries is highly scarce. We therefore consider a natural direction: synthesizing training data with LLMs~\cite{majumdar2025genetic, adarsh2025siked}.
However, naively prompting LLMs to directly generate such training data about private libraries faces some critical challenges. 
To illustrate this, we conduct an additional empirical study on \texttt{ndonnx}, a library that is unfamiliar to all evaluated models in this paper.
Specifically, we inject the complete API documentation of the library into the prompt and ask DeepSeek-V3-671B to directly synthesize $200$ training instances, each consisting of a coding requirement and its corresponding solution.

Our analysis reveals two major challenges that prevent such naively synthesized data from being directly used for training:
\ding{182} \bi{Low Data Diversity.} We find that 74\% of the synthesized samples invoke fewer than four APIs. This suggests that when LLMs are unfamiliar with the target private library, they tend to generate overly basic requirements, with many samples relying on only a small set of APIs. In contrast, real-world private-library-oriented development involves substantially more diverse and multiple API compositions. As a result, naively synthesized data exhibits limited diversity and fails to provide broad coverage of realistic development needs.
\ding{183} \bi{Poor Data Quality.} Through manual inspection, we find that 46\% of the synthesized samples contain obvious runtime errors, often caused by invoking non-existent APIs or using existing APIs incorrectly. More importantly, this only reflects failures that are immediately observable at execution time. Many additional samples may still suffer from functional errors even if they can run successfully. Such low-quality data is unreliable for training and may even be harmful if used directly for fine-tuning.

\begin{boxK}
\small \faIcon{pencil-alt} \textbf{Motivation 2:}
Directly prompting LLMs to synthesize training data for private libraries faces two major challenges, namely \emph{Low Data Diversity} and \emph{Poor Data Quality}.
\end{boxK}

Building on the above motivation, we propose \emph{Progressive Graph Evolution} and \emph{Multidimensional Graph Pruning} in the next section, to automatically synthesize high-diversity and high-quality training data for private-library-oriented code generation.

\section{Methodology}
\label{sec:method}

In this section, we propose \name, an approach designed to teach LLMs how to \emph{invoke} private-library APIs effectively.

\subsection{Overview}
\label{sec:method-1}

At a high level, \name first synthesizes training data tailored to the target private library and then fine-tunes the model on the synthesized data.
Algorithm~\ref{alg:overview} summarizes the overall procedure.
\seclrevision{\name uses a \bi{synthesis graph} as a presentation device to illustrate how training samples are generated from API specifications or earlier samples.} Formally, let $\mathcal{L}$ denote a private library with API set $\mathcal{A}=\{a_1,\dots,a_{|\mathcal{A}|}\}$ and corresponding specifications $\mathcal{S}=\{s(a)\mid a\in\mathcal{A}\}$, where $s(a)$ contains the signature and functional description of API $a$. Based on $\mathcal{L}$, we construct a synthesis graph $\mathcal{G}=(\mathcal{V},\mathcal{E})$, whose nodes consist of \bi{API nodes} and \bi{sample nodes}. Each API node corresponds to one private-library API specification $s(a)$, while each sample node corresponds to one synthesized training sample $(r,y)$, where $r$ is a coding requirement and $y$ is its reference solution. An edge $(u,v)\in\mathcal{E}$ indicates that node $v$ is synthesized from node $u$. Since each new sample node is generated only from pre-existing nodes, $\mathcal{G}$ is naturally a Directed Acyclic Graph (DAG).

\begin{algorithm}[!htbp]
\caption{Overall Procedure of \name}
\label{alg:overview}
\small
\begin{algorithmic}[1]
\Require Private-library API specifications $\mathcal{S}$, LLM $p_{\theta}$, Target dataset size $N$
\Ensure Fine-tuned LLM $p_{\hat{\theta}}$

\Procedure{\name}{$\mathcal{S}, p_{\theta}, N$}
    \State Initialize synthesis graph $\mathcal{G} = (\mathcal{V}, \mathcal{E})$ with $\mathcal{V} \gets \mathcal{S}, \mathcal{E} \gets \emptyset$
    
    \While{$|\mathcal{V} \setminus \mathcal{S}| < N$}
        \State $\mathcal{G}, v_{\text{new}} \gets \Call{Evolution}{\mathcal{G}, \mathcal{S}}$  \Comment{\textit{Progressive Graph Evolution}}
        \State $\mathcal{G} \gets \Call{Pruning}{\mathcal{G}, v_{\text{new}}}$ \Comment{\textit{Multidimensional Graph Pruning}}
    \EndWhile
    
    \State $\mathcal{D}_{\text{syn}} \gets \{(r, y) \mid (r, y) \in \mathcal{V} \setminus \mathcal{S}\}$
    \State Fine-tune $p_{\theta}$ on $\mathcal{D}_{\text{syn}}$ optimizing Eq.~\ref{eq:sft} to obtain $p_{\hat{\theta}}$ \Comment{\textit{Training}}
    
    \State \Return $p_{\hat{\theta}}$
\EndProcedure
\end{algorithmic}
\end{algorithm}

\begin{figure*}[!tbp]
    \centering
    \includegraphics[width=0.88\textwidth]{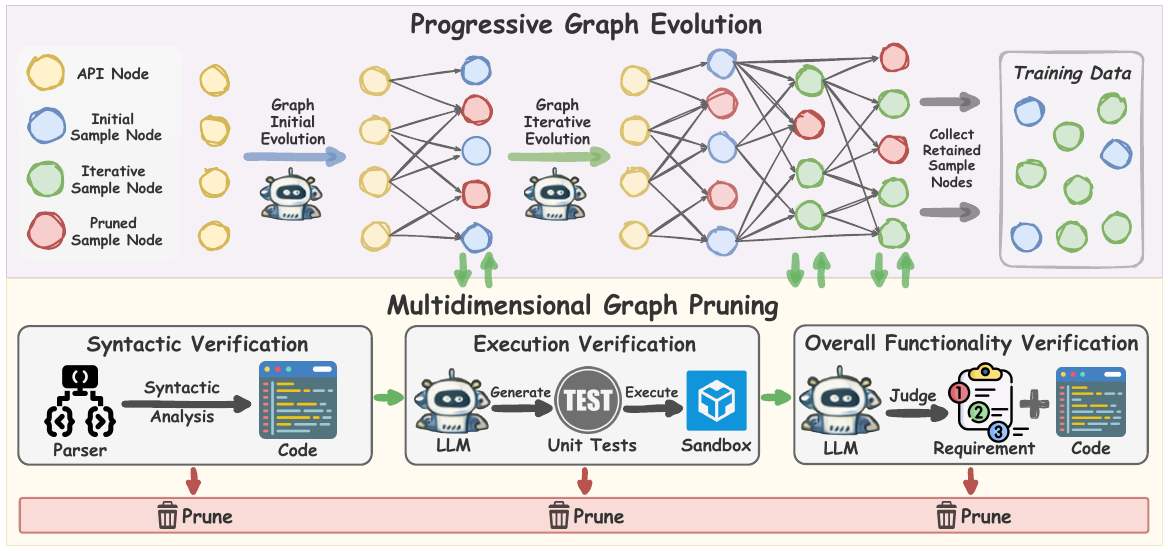}

    \caption{\lrevision{Overview of \name. Starting from API nodes, Progressive Graph Evolution constructs increasingly diverse sample nodes, while Multidimensional Graph Pruning removes low-quality nodes through syntactic verification, execution verification, and overall functionality verification.}}
    \label{fig:overview}

\end{figure*}

\lrevision{Starting from API nodes, \name progressively constructs and filters candidate training samples.} As discussed in Section~\ref{sec:motivation}, naively synthesized data often suffers from low diversity and poor quality. \lrevision{To address these challenges, \name combines two synthesis operators. First, \bi{Progressive Graph Evolution} (Section~\ref{sec:method-2}) synthesizes new sample nodes from existing nodes, moving from isolated API-level knowledge toward increasingly diverse API-composition scenarios. Second, \bi{Multidimensional Graph Pruning} (Section~\ref{sec:method-3}) filters low-quality sample nodes and retains only samples that pass automatic verification.} \lrevision{The retained sample nodes form the synthetic dataset $\mathcal{D}_{\text{syn}}=\{(r_i,y_i)\}_{i=1}^{N}$.} Figure~\ref{fig:overview} illustrates the overall workflow.

We then use $\mathcal{D}_{\text{syn}}$ to fine-tune a model $p_{\theta}$ for private-library-oriented code generation (Section~\ref{sec:method-4}).
Given a coding requirement $r$, the model is trained to generate a functionally correct solution $y$ by optimizing the standard maximum-likelihood objective:
\begin{equation}
\mathcal{L}(\theta)
= - \sum_{(r,y)\in \mathcal{D}_{\text{syn}}} \log p_{\theta}(y \mid r),
\qquad
\hat{\theta}=\arg\min_{\theta}\mathcal{L}(\theta).
\label{eq:sft}
\end{equation}

\subsection{Progressive Graph Evolution}
\label{sec:method-2}

To address the low-diversity challenge identified in Section~\ref{sec:motivation}, we introduce \emph{Progressive Graph Evolution} (Algorithm~\ref{alg:evolution}), \lrevision{which synthesizes API-use examples in two stages. It first generates \textit{initial sample nodes} from \textit{API nodes} and then produces \textit{iterative sample nodes} from existing sample nodes.} \lrevision{Through these two stages, \name expands the synthesized data from basic API use to more diverse API compositions.}

\begin{algorithm}[!tbp]
\caption{Progressive Graph Evolution}
\label{alg:evolution}
\small
\begin{algorithmic}[1]
\Require Current graph $\mathcal{G} = (\mathcal{V}, \mathcal{E})$, API specifications $\mathcal{S}$
\Ensure Updated graph $\mathcal{G}$ and the newly added sample node $v_{\text{new}}$

\Procedure{Evolution}{$\mathcal{G}, \mathcal{S}$}
    \If{$\mathcal{G}$ lacks sufficient initial sample nodes} \Comment{\textit{Graph Initial Evolution}}
        \State Sample $\mathcal{S}_{\text{api}} \subset \mathcal{S}$
        \State $(r, y) \gets \mathrm{Evolve}(\mathcal{S}_{\text{api}}, \mathcal{I}_{\text{init}})$
        \State $\mathcal{V}_{\text{parents}} \gets \mathcal{S}_{\text{api}}$
    \Else \Comment{\textit{Graph Iterative Evolution}}
        \State Sample $S_{\text{sample}} \subset \mathcal{V} \setminus \mathcal{S}$
        \State $(r, y) \gets \mathrm{Evolve}(S_{\text{sample}}, \mathcal{I}_{\text{iter}})$
        \State $\mathcal{V}_{\text{parents}} \gets S_{\text{sample}}$
    \EndIf
    
    \State $v_{\text{new}} \gets (r, y)$
    \State $\mathcal{V} \gets \mathcal{V} \cup \{v_{\text{new}}\}$
    \State $\mathcal{E} \gets \mathcal{E} \cup \{(u, v_{\text{new}}) \mid u \in \mathcal{V}_{\text{parents}}\}$
    
    \State \Return $\mathcal{G}, v_{\text{new}}$
\EndProcedure
\end{algorithmic}
\end{algorithm}

\bi{(1) Graph Initial Evolution.}
At the beginning, the graph contains only API nodes.
The goal of graph initial evolution is to expose the model to diverse and comprehensive private-API knowledge before more complex samples are synthesized.
To encourage broader coverage of the target library, we adopt a dynamic long-tail-biased sampling strategy. Specifically, let $c_i$ denote the number of accepted sample nodes that invoke API $a_i$. We assign each API $a_i$ a sampling weight:
\begin{equation}
w_i = \frac{1}{(c_i+1)^\alpha}.
\label{eq:api-sampling-weight}
\end{equation}
where $\alpha$ controls the bias strength. We then sample APIs without replacement according to these weights and update $\{c_i\}$ whenever a synthesized sample is accepted, so that APIs covered less frequently are more likely to be selected later. We use the associated specifications $\mathcal{S}_{\text{api}}$ as seeds.
We then prompt the LLM to synthesize a sample node based on these seed APIs:
\begin{equation}
(r, y) = \mathrm{Evolve}(\mathcal{S}_{\text{api}}, \mathcal{I}_{\text{init}}),
\label{eq:evolve-api}
\end{equation}
where $\mathcal{I}_{\text{init}}$ denotes the instruction prompt for graph initial evolution, with its template shown in the Supplementary Materials; $r$ is the generated coding requirement, and $y$ is the corresponding reference solution. 
This process produces the initial sample nodes in the graph. Since randomly sampled APIs may not always be tightly related, we do not strictly require the LLM to invoke every seed API in the generated solution.

\bi{(2) Graph Iterative Evolution.}
\lrevision{Once a pool of sample nodes has been obtained, we iteratively synthesize new sample nodes from existing ones.} Specifically, we randomly sample a small set of existing sample nodes $S_{\text{sample}}=\{(r_i,y_i)\}_{i=1}^{m}$ and prompt the LLM to evolve them into a new sample using $\mathcal{I}_{\text{iter}}$, whose template is provided in the Supplementary Materials:
\begin{equation}
(r', y') = \mathrm{Evolve}(S_{\text{sample}}, \mathcal{I}_{\text{iter}}),
\label{eq:evolve-sample}
\end{equation}
where $\mathcal{I}_{\text{iter}}$ is the instruction prompt for graph iterative evolution. The operator $\mathrm{Evolve}(\cdot)$ instructs the LLM to integrate multiple existing requirements into a single coherent and more diverse requirement $r'$, and to produce a reference solution $y'$ that satisfies this new requirement through coordinated invocation of multiple private-library APIs.
\lrevision{Repeatedly applying Graph Iterative Evolution yields samples that combine increasingly varied API-use patterns.}

\begin{algorithm}[!tbp]
\caption{Multidimensional Graph Pruning}
\label{alg:pruning}
\small
\begin{algorithmic}[1]
\Require Current graph $\mathcal{G} = (\mathcal{V}, \mathcal{E})$, Newly added node $v_{\text{new}} = (r, y)$
\Ensure Pruned graph $\mathcal{G}$

\Procedure{Pruning}{$\mathcal{G}, v_{\text{new}}$}
    \State $V_{\text{syn}} \gets \text{Parse } y \text{ using standard AST parser}$ \Comment{\textit{Syntactic Verification}}
    \State $V_{\text{exe}} \gets \text{Execute } y \text{ against generated unit tests}$ \Comment{\textit{Execution Verification}}
    \State $V_{\text{fun}} \gets \text{LLM-as-a-Judge assesses } (r, y)$ \Comment{\textit{Overall Functionality Verification}}
    
    \If{$V_{\text{syn}} \cdot V_{\text{exe}} \cdot V_{\text{fun}} == 0$}
        \State $\mathcal{V} \gets \mathcal{V} \setminus \{v_{\text{new}}\}$
        \State $\mathcal{E} \gets \mathcal{E} \setminus \{(u, v_{\text{new}}) \in \mathcal{E}\}$
    \EndIf
    
    \State \Return $\mathcal{G}$
\EndProcedure
\end{algorithmic}
\end{algorithm}

\subsection{Multidimensional Graph Pruning}
\label{sec:method-3}

To address the poor-quality challenge identified in Section~\ref{sec:motivation}, we introduce \emph{Multidimensional Graph Pruning} (Algorithm~\ref{alg:pruning}), \lrevision{which filters low-quality sample nodes through multidimensional verification.}
For any new sample node $(r,y)$, we retain it in the graph only if it passes three complementary checks that target different failure modes: syntactic correctness, execution validity, and overall functionality. Formally, we define three binary validators $V_{\text{syn}}, V_{\text{exe}}, V_{\text{fun}}\in\{0,1\}$ and retain a sample node in the graph only if
\begin{equation}
V(r,y)=V_{\text{syn}}(r,y)\cdot V_{\text{exe}}(r,y)\cdot V_{\text{fun}}(r,y)=1.
\label{eq:verify}
\end{equation}
\lrevision{We apply these validators immediately after a candidate node is created and before it can be selected as a parent in subsequent evolution. This eager pruning prevents rejected nodes from propagating errors to later sample nodes.}

\bi{(1) Syntactic Verification.}
We first perform lightweight static verification to ensure the basic well-formedness of the generated code. Specifically, we parse the reference solution $y$ using the standard Abstract Syntax Tree (AST) parser of the target programming language. Any sample that fails to parse is discarded. This step removes obviously invalid code before invoking more expensive verification procedures.

\bi{(2) Execution Verification.}
Syntactic correctness does not imply valid private-API invocation. To better identify invocation errors, we verify executability through test-based execution. For each sample $(r,y)$, we prompt the LLM to generate a set of unit tests $T=\{t_j\}$ using the template in the Supplementary Materials, including representative inputs and corresponding assertions. We then execute $y$ against $T$ in a sandboxed environment. Samples are discarded if they raise runtime errors or fail any assertion.

\bi{(3) Overall Functionality Verification.}
Even if a sample is executable, the synthesized requirement and solution may still be unrealistic or semantically misaligned.
To further improve data quality, we follow the widely adopted LLM-as-a-Judge paradigm~\cite{he2026llm, feng2025we, do2025generate} to assess the overall functionality of the synthesized samples.
Concretely, we prompt a judge LLM with the template in the Supplementary Materials to assess each synthesized sample along two dimensions: (i) whether the requirement $r$ is realistic and well-defined, and (ii) whether the solution $y$ satisfies the intent of $r$. Any sample that fails this evaluation is also discarded.

\subsection{Training and Deployment}
\label{sec:method-4}

After repeatedly applying Progressive Graph Evolution and Multidimensional Graph Pruning, we collect all retained sample nodes in the final graph and construct the synthetic training set $\mathcal{D}_{\text{syn}}$. We then fine-tune the model on $\mathcal{D}_{\text{syn}}$ using the objective in Eq.~\ref{eq:sft}.

After training, the model can be directly applied to private-library-oriented code generation. Given a coding requirement $r$, the deployed model generates code as:
\begin{equation}
\hat{y}=\arg\max_{y} p_{\hat{\theta}}(y\mid r),
\label{eq:deploy_plain}
\end{equation}
where the learned parameters $\hat{\theta}$ encode the model's learned ability to invoke private-library APIs.

Moreover, we view \name as orthogonal to prior RAG-based approaches. Existing approaches primarily improve how well the model can \textit{see} the right API information at inference time, whereas \name focuses on enabling the model to learn how to \textit{invoke} private-library APIs effectively. In practice, the two directions can be combined. Let $\mathrm{Retrieve}(r)$ return relevant API knowledge. We can condition the fine-tuned model on the augmented prompt:
\begin{equation}
\hat{y}=\arg\max_{y} p_{\hat{\theta}}\big(y\mid [r;\mathrm{Retrieve}(r)]\big),
\label{eq:deploy_rag}
\end{equation}
where $[\,\cdot\,;\,\cdot\,]$ denotes concatenation.

\section{Benchmark Construction}
\label{sec:benchmark-construct}

Evaluating private-library-oriented code generation requires target libraries unseen by the tested models. However, many existing benchmarks are built on libraries released long ago and may already be familiar to recent LLMs.
For example, \dataname{TorchDataEval}~\cite{apifinder}, a benchmark widely used in prior work~\cite{apifinder, zan2024diffcoder, zan2025private}, evaluates models on the TorchData library~\cite{torchdata}, which was released in December 2021, well before the knowledge cutoffs of most modern LLMs.
Therefore, gains on such benchmarks do not necessarily provide convincing evidence of improved private-library-oriented code generation.

To enable a more rigorous evaluation, we construct two benchmarks, \textbf{\ndonnx} and \textbf{\numba}.
These benchmarks are designed to better reflect realistic private-library-oriented code generation scenarios, in which the target libraries are largely absent from the training corpora of modern LLMs.
We describe the construction process below.

\bi{Library Selection.}
The primary requirement is to choose libraries that are unfamiliar to the evaluated models while still being realistic targets for developers.
We therefore screened candidate libraries according to four criteria.
\ding{182} First, the library should be recent, so that its APIs are unlikely to have been extensively observed during model pre-training.
\ding{183} Second, it should expose a non-trivial API surface, making it possible to construct tasks that require API composition rather than simple single-call invocation.
\ding{184} Third, the library should be practically meaningful and executable in a standard experimental environment, so that benchmark instances reflect realistic programming requirements instead of artificial API exercises.
\ding{185} Fourth, the library should be actively maintained and accompanied by usable documentation, which is necessary for reliable annotation and later replication.
Following these criteria, we select two libraries, \texttt{ndonnx}~\cite{ndonnx} and \texttt{numba-cuda}~\cite{numba}.
Both libraries were first released in 2024 and underwent substantial development throughout 2025 and 2026, making them suitable for evaluating private-library-oriented code generation under limited prior model exposure.
They also cover different application domains: \texttt{ndonnx} focuses on array-oriented ONNX model construction, whereas \texttt{numba-cuda} targets CUDA programming in Python.
This diversity allows the benchmarks to evaluate whether a method can generalize across distinct styles of private-library usage.

\bi{Instance Construction.}
For each library, five annotators with extensive programming experience (more than four years on average) manually constructed benchmark instances.
Each instance consists of a natural-language coding requirement and a set of unit tests used for evaluation.
\ding{182} The annotators first studied the official documentation, API references, and examples of the target library to understand its intended usage patterns.
\ding{183} They then designed requirements that resemble realistic developer requests, such as transforming data structures, constructing computational graphs, configuring execution behavior, or combining multiple library primitives to implement a higher-level operation.
To avoid overly shallow instances, annotators were instructed to prioritize tasks that require coordinated use of multiple APIs, while excluding requirements that can be solved by a single obvious call or by using only standard Python libraries.
They also avoided directly copying documentation examples, so that the resulting instances test API understanding and adaptation rather than memorization.
\ding{184} For each requirement, annotators wrote unit tests to check functional correctness.
To improve the efficiency of test construction, annotators were allowed to use GPT-5.2~\cite{openai2025gpt52} to draft candidate test cases, but all tests were manually reviewed and revised before being included in the benchmark.

\bi{Verification and Refinement.}
To ensure benchmark quality, all constructed instances went through multiple rounds of manual verification and refinement.
\ding{182} Each requirement was independently reviewed by two annotators, who checked its clarity, completeness, and realism, while filtering out instances that were duplicated.
\ding{183} Each test case was also independently reviewed by two annotators to remove ambiguous assertions and redundant cases.
\ding{184} For every retained requirement, one annotator implemented a reference solution and executed it against the corresponding test suite to verify solvability.
An instance was discarded if it failed any of the above verification steps.
\lrevision{For each initial instance, two annotators independently checked both the requirement and its test cases.
To assess the consistency of this review process, we calculate agreement as the proportion of initial instances for which the two annotators reached the same judgment on both components.
The agreement rates are 92.33\% for \ndonnx and 87.00\% for \numba, indicating that the independent reviews were generally consistent across both benchmarks.}

\begin{table*}[!tbp]
\centering
\small
\caption{Overview of the constructed benchmarks. \#Instances denotes the number of instances in each benchmark; Avg.\ APIs denotes the average number of required APIs per instance; Avg.\ Tests denotes the average number of test cases per instance; Library denotes the target private library; First Release denotes the initial release time of the target library; Used Version Date denotes the release time of the library version used in this paper; and \#APIs denotes the total number of APIs in the target library.} 
\label{tab:benchmark-stats}

\begin{adjustbox}{max width=\linewidth}
\begin{tabular}{lccccccc}
\toprule
\textbf{Benchmark} & \textbf{\#Instances} & \textbf{Avg.\ APIs} & \textbf{Avg.\ Tests} & \textbf{Library} & \textbf{First Release} & \textbf{Used Version Date} & \textbf{\#APIs} \\
\midrule
\ndonnx & 169 & 4.56 & 9.00 & \texttt{\href{https://pypi.org/project/ndonnx/0.17.1/}{ndonnx}} & 2024.06 & 2025.12 & 179 \\
\numba  & 187 & 6.73 & 9.10 & \texttt{\href{https://pypi.org/project/numba-cuda/0.27.0/}{numba-cuda}} & 2024.06 & 2026.02 & 725 \\
\bottomrule
\end{tabular}
\end{adjustbox}

\end{table*}

After more than 150 person-hours of collective effort, we finalized 169 instances for \ndonnx and 187 instances for \numba.
On average, each instance contains approximately 9 test cases and requires the coordinated use of more than 4 distinct APIs.
Detailed statistics of the two benchmarks are summarized in Table~\ref{tab:benchmark-stats}. 
To facilitate future research, we have made the complete benchmarks publicly available at \url{https://github.com/THU-Agent/PriCoder/tree/main/data/benchmark}.

\section{Experimental Setup}
\label{sec:setup}

To assess \name, we conduct comprehensive experiments to answer four Research Questions (RQs). In this section, we present the details of our experimental setup, including benchmarks, evaluation metrics, baselines, models, and other implementation details.

\subsection{Research Questions}
\label{sec:rq}

Our study aims to answer the following RQs.

\textit{RQ1: How does \name perform in private-library-oriented code generation?}
This RQ aims to verify that \name can effectively teach LLMs to invoke private-library APIs, thereby achieving superior results in private-library-oriented tasks.
To answer this RQ, we compare \name with multiple baselines across three LLMs.

\textit{RQ2: Does \name negatively affect general code generation capability?}
Updating model parameters may introduce unintended degradation on general-purpose coding tasks. To assess this risk, we evaluate \name on widely used public code generation benchmarks that do not involve the private libraries.

\textit{RQ3: What is the contribution of each component in \name?}
Since \name consists of two key components, we conduct ablation studies to isolate their individual effects and quantify their contributions to overall performance.

\textit{RQ4: How do key factors affect the effectiveness of \name?}
This RQ studies how key factors influence the effectiveness of \name. \lrevision{In particular, we focus on four key factors: the scale of the synthesized training data, the LLM used for data synthesis, the allocation between Graph Initial Evolution and Graph Iterative Evolution, and the API sampling bias strength $\alpha$.}

\subsection{Benchmarks}
\label{sec:benchmark}
We evaluate \name on \lrevision{four} benchmarks, including two novel private-library benchmarks and \lrevision{two} widely used public benchmarks.

\textbf{\numba} and \textbf{\ndonnx} (Section~\ref{sec:benchmark-construct}). We introduce two novel benchmarks for rigorous evaluation of private-library-oriented code generation. \numba is built on the \texttt{numba-cuda} library, while \ndonnx is built on the \texttt{ndonnx} library.

\textbf{\humaneval}~\cite{humaneval}. We include this existing benchmark to assess general code generation capability. It consists of 164 hand-written programming tasks and mainly involves built-in functions, without requiring any third-party libraries.

\lrevision{\textbf{\mbpp}~\cite{austin2021program}. We also include \mbpp, a benchmark of short Python programming problems specified in natural language. The evaluation subset used in our experiments contains 378 tasks and does not require the target private libraries.}

\subsection{Metrics}
\label{sec:metric}

We use \pass{k} and \exec{k} as our main evaluation metrics. To reduce the randomness and obtain more reliable estimates, we compute both metrics using the standard unbiased estimator.

\textbf{\pass{k}.}
For each instance, we sample $n \ge k$ candidate solutions (we use $n=10$ and $k\in\{1,3,5\}$), execute the provided test cases, and count the number of passing solutions $c$. Following prior work~\cite{chen2022codet, athiwaratkun2022multi}, we compute \pass{k} using the unbiased estimator:
\begin{equation}
\pass{k}=\mathbb{E}_{\text{instances}}\!\left[1-\frac{\binom{n-c}{k}}{\binom{n}{k}}\right].
\end{equation}

\textbf{\exec{k}.}
As shown in Section~\ref{sec:motivation}, models often misuse private APIs and trigger runtime failures. We therefore report \exec{k} to measure basic executability. It is defined analogously to \pass{k}, except that a solution is counted as successful if it runs to completion on the test inputs without raising any exception.

\subsection{Baselines}
\label{sec:baseline}

Our core contribution lies in enabling models to automatically learn how to \emph{use} private libraries. Therefore, our fundamental baseline is the original model that has not been applied with \name, which we denote as \textbf{\baselinename{Vanilla}}.

Most existing approaches for private-library-oriented code generation are based on RAG. Although they are orthogonal to \name, we include three representative approaches for comparison.

\begin{itemize}[leftmargin=12pt]
\item \textbf{\baselinename{Naive RAG}~\cite{docprompting, apifinder}.} It adopts a retrieval-augmented generation pipeline to search for relevant APIs and include their specifications directly in the prompt.
\item \textbf{\baselinename{EpiGen}~\cite{epigen}.} It utilizes an LLM to decompose complex coding requirements into subtasks using the prompt in the Supplementary Materials, and retrieves APIs for each subtask to improve the relevance of injected knowledge.
\item \textbf{\baselinename{CAPIR}~\cite{capir}.} Building on task decomposition with the same prompt in the Supplementary Materials, \baselinename{CAPIR} further reranks retrieved APIs using the prompt in the Supplementary Materials to improve the accuracy of injected API knowledge.
\end{itemize}

\lrevision{Furthermore, we include a baseline that represents an upper bound for documentation-based RAG approaches that use API specifications as inference-time context.} In this setting, we explicitly provide the LLM with the specifications of all APIs required to solve the specific coding requirements. Consistent with Section~\ref{sec:motivation}, we term this setting \textbf{\baselinename{Oracle}}.

\subsection{Models}
\label{sec:model}

We apply \name to three widely used LLMs that lack prior training exposure to the target libraries in \numba and \ndonnx: DeepSeek-Coder-6.7B-Instruct~\cite{deepseekcoder}, Qwen2.5-Coder-7B-Instruct~\cite{qwencoder}, and LLaMA3.1-8B-Instruct~\cite{llama}.
For brevity, we hereafter refer to them as \deepseek, \qwen, and \llama.

Due to computational constraints, our main experiments focus on models with fewer than 20B parameters, and extending \name to larger models is left for future work.
However, we believe this setting is still representative and applicable in real-world scenarios.
In many enterprise scenarios, private libraries may contain sensitive or proprietary information, making it difficult to access large state-of-the-art models through external APIs.
Therefore, focusing on locally deployable models with moderate sizes aligns with a practical deployment setting for private-library-oriented code generation.

\subsection{Implementation Details}
\label{sec:implementation}

For \name, unless otherwise specified, we use each target model to synthesize its own training data to avoid confounding the results with potential distillation from a stronger model.
For \texttt{numba-cuda}, we synthesize 6K training instances per model, and for \texttt{ndonnx}, we synthesize 20K training instances per model. The two synthesized datasets are used to train separate models for their corresponding benchmarks. 
During data synthesis, the number of samples produced by Graph Initial Evolution and Graph Iterative Evolution is controlled at a ratio of $1:2$.
In Graph Initial Evolution, we sample 1--5 API nodes as synthesis seeds.
In Graph Iterative Evolution, we sample 2--3 existing sample nodes and prompt the model to evolve them into a new sample.
For the long-tail-biased API sampling in Graph Initial Evolution, we set the bias strength $\alpha$ to $0.7$.
In Execution Verification, we require the model to generate at least three test cases for each synthesized sample.
We fine-tune the models using LoRA~\cite{hu2022lora} with AdamW~\cite{loshchilov2017decoupled} for one epoch and a batch size of 16.
The LoRA rank is set to 16, the LoRA alpha is set to 32, and the learning rate is set to $1.0\times10^{-4}$.

For baselines that require task decomposition or reranking, we use the same LLM as the backbone for these steps. We adopt the widely used BGE-M3~\cite{multi2024m3} as the embedding model for retrieval. 
For baseline-specific hyperparameters, we perform grid search on \ndonnx and use the best-performing configuration.
For \baselinename{Naive RAG}, the number of retrieved APIs is set to 10.
For \baselinename{EpiGen}, we retrieve 4 APIs for each decomposed subtask.
For \baselinename{CAPIR}, we retrieve 6 APIs for each subtask before reranking.

To ensure fair comparison, all methods, including \name, generate 10 samples per requirement with identical decoding settings, using temperature 0.5, top-$p$ 0.95, and a maximum of 800 new tokens. 
We use vLLM~\cite{kwon2023efficient} as the inference framework and LLaMA-Factory~\cite{zheng2024llamafactory} for model fine-tuning. All experiments are conducted on a server equipped with eight NVIDIA A100 GPUs, each with 80 GB of memory.

\section{Experimental Results}
\label{sec:exp}

\subsection{RQ1: How does \name perform in private-library-oriented code generation?}
\label{sec:rq1}

The primary objective of \name is to improve models' ability to correctly invoke private-library APIs. 
In this RQ, we evaluate whether this learned invocation ability translates into better performance on private-library-oriented code generation tasks.

\vspace{2pt}
\noindent \textbf{Setting.}
We apply all baselines and \name to the three models described in Section~\ref{sec:model}, and evaluate them on the two benchmarks introduced in Section~\ref{sec:benchmark-construct}. 
We report both \pass{k} and \exec{k}, where $k\in\{1,3,5\}$.
As discussed in Section~\ref{sec:method-4}, \name is orthogonal to existing RAG-based approaches: RAG provides API knowledge at inference time, whereas \name teaches the model how to invoke private-library APIs through training.
Therefore, in addition to evaluating each method independently, we further combine \name with the RAG-based baselines following Eq.~\ref{eq:deploy_rag}.

\vspace{2pt}
\noindent \textbf{Results.}
Table~\ref{tab:rq1} reports the results of all methods on \ndonnx and \numba.

\begin{table*}[!tbp]
\centering
\small
\renewcommand{\arraystretch}{1.02}
\setlength{\tabcolsep}{2pt}
\caption{\pass{k} and \exec{k} (\%) on \ndonnx and \numba.
}
\label{tab:rq1}

\begin{adjustbox}{max width=\linewidth}
\begin{tabular}{c|l|cccccc|cccccc}
\toprule
\multirow{2}{*}{\textbf{Model}} & \multirow{2}{*}{\textbf{Method}} &
\multicolumn{6}{c|}{\textbf{\ndonnx}} &
\multicolumn{6}{c}{\textbf{\numba}} \\
\cmidrule(lr){3-8}\cmidrule(lr){9-14}
& & {\footnotesize\bfseries \pass{1}} & {\footnotesize\bfseries \pass{3}} & {\footnotesize\bfseries \pass{5}} & {\footnotesize\bfseries \exec{1}} & {\footnotesize\bfseries \exec{3}} & {\footnotesize\bfseries \exec{5}}
  & {\footnotesize\bfseries \pass{1}} & {\footnotesize\bfseries \pass{3}} & {\footnotesize\bfseries \pass{5}} & {\footnotesize\bfseries \exec{1}} & {\footnotesize\bfseries \exec{3}} & {\footnotesize\bfseries \exec{5}} \\
\midrule

\multirow{10}{*}{\shortstack{\fontfamily{pcr}\selectfont DeepSeek\\\fontfamily{pcr}\selectfont-6.7B}}
& \baselinename{Vanilla} 
& 20.36 & 29.16 & 33.26 & 25.74 & 37.09 & 42.20
& 10.59 & 21.26 & 27.82 & 22.25 & 42.34 & 52.23  \\
\rowcolor{namerow}\cellcolor{white} 
& \ +\name                  
& 48.17 & 60.36 & 64.92 & 61.72 & 74.19 & 78.58 
& 24.12 & 38.47 & 45.04 & 52.73 & 74.26 & 82.21  \\
[-0.3ex]\arrayrulecolor[gray]{0.5}\hhline{~|-------------}\arrayrulecolor{black}
& \baselinename{Naive RAG} 
& 14.97 & 25.16 & 29.87 & 25.15 & 38.75 & 43.85 
&  8.07 & 18.05 & 24.94 & 17.38 & 38.04 & 50.71  \\
\rowcolor{namerow}\cellcolor{white} & \ +\name                  
& 46.80 & 60.60 & 65.60 & 62.49 & 76.03 & 80.58 
& 19.30 & 32.66 & 39.01 & 43.48 & 66.74 & 75.76  \\
[-0.3ex]\arrayrulecolor[gray]{0.5}\hhline{~|-------------}\arrayrulecolor{black}
& \baselinename{EpiGen} 
& 16.86 & 25.78 & 30.27 & 27.04 & 40.32 & 46.31 
& 11.18 & 23.61 & 30.97 & 22.83 & 45.91 & 57.40  \\
& \cellcolor{namerow} \ +\name                  
& \cellcolor{namerow} 44.79 & \cellcolor{namerow} 59.04 & \cellcolor{namerow} 63.39 & \cellcolor{namerow} 61.89 & \cellcolor{namerow} 74.95 & \cellcolor{namerow} 78.95 
& \cellcolor{namerow} 24.49 & \cellcolor{namerow} 38.70 & \cellcolor{namerow} 45.28 & \cellcolor{namerow} 51.39 & \cellcolor{namerow} 72.05 & \cellcolor{namerow} 79.73  \\
[-0.3ex]\arrayrulecolor[gray]{0.5}\hhline{~|-------------}\arrayrulecolor{black}
& \baselinename{CAPIR} 
& 18.76 & 29.22 & 34.15 & 28.88 & 41.27 & 45.72 
& 10.32 & 21.52 & 27.82 & 20.00 & 40.64 & 51.32  \\
\rowcolor{namerow}\cellcolor{white} & \ +\name                  
& 47.10 & 60.01 & 65.04 & 62.37 & 75.90 & 80.57 
& 24.28 & 38.30 & 44.81 & 53.85 & 74.81 & 81.54  \\
[-0.3ex]\arrayrulecolor[gray]{0.5}\hhline{~|-------------}\arrayrulecolor{black}
& \baselinename{Oracle} 
& 38.05 & 50.82 & 55.86 & 51.01 & 67.34 & 72.96 
& 17.97 & 34.23 & 42.45 & 30.32 & 55.73 & 66.82  \\
\rowcolor{namerow}\cellcolor{white} & \ +\name                  
& 55.44 & 67.56 & 71.79 & 71.01 & 82.86 & 86.94 
& 28.66 & 43.24 & 48.88 & 55.19 & 77.48 & 84.32  \\
\midrule

\multirow{10}{*}{\shortstack{\fontfamily{pcr}\selectfont Qwen\\\fontfamily{pcr}\selectfont-7B}}
& \baselinename{Vanilla} 
& 23.37 & 31.69 & 35.42 & 28.52 & 38.91 & 43.40 
& 16.04 & 31.82 & 39.51 & 26.74 & 51.56 & 63.08  \\
\rowcolor{namerow}\cellcolor{white} & \ +\name                  
& 47.99 & 59.25 & 62.76 & 55.86 & 67.58 & 71.68 
& 35.61 & 51.83 & 58.29 & 56.47 & 77.54 & 83.60  \\
[-0.3ex]\arrayrulecolor[gray]{0.5}\hhline{~|-------------}\arrayrulecolor{black}
& \baselinename{Naive RAG} 
& 28.40 & 35.89 & 38.52 & 37.04 & 44.69 & 47.53 
& 16.68 & 33.18 & 41.55 & 26.63 & 50.44 & 61.57  \\
\rowcolor{namerow}\cellcolor{white} & \ +\name                  
& 45.38 & 55.74 & 59.43 & 56.57 & 67.31 & 71.09 
& 33.37 & 47.34 & 53.28 & 55.78 & 73.64 & 79.56  \\
[-0.3ex]\arrayrulecolor[gray]{0.5}\hhline{~|-------------}\arrayrulecolor{black}
& \baselinename{EpiGen} 
& 28.82 & 35.69 & 38.00 & 33.73 & 41.48 & 44.46 
& 16.47 & 32.75 & 40.70 & 27.43 & 52.33 & 63.15  \\
& \cellcolor{namerow} \ +\name                  
& \cellcolor{namerow} 47.63 & \cellcolor{namerow} 56.49 & \cellcolor{namerow} 59.58 & \cellcolor{namerow} 57.93 & \cellcolor{namerow} 68.09 & \cellcolor{namerow} 71.50 
& \cellcolor{namerow} 35.08 & \cellcolor{namerow} 49.74 & \cellcolor{namerow} 56.69 & \cellcolor{namerow} 56.84 & \cellcolor{namerow} 75.43 & \cellcolor{namerow} 81.06  \\
[-0.3ex]\arrayrulecolor[gray]{0.5}\hhline{~|-------------}\arrayrulecolor{black}
& \baselinename{CAPIR} 
& 29.41 & 36.82 & 39.68 & 36.21 & 43.68 & 46.79 
& 14.81 & 29.16 & 36.76 & 25.45 & 48.45 & 59.92  \\
\rowcolor{namerow}\cellcolor{white} & \ +\name                  
& 44.44 & 54.66 & 58.58 & 56.92 & 67.57 & 70.97 
& 32.46 & 48.09 & 54.49 & 52.89 & 73.09 & 79.57  \\
[-0.3ex]\arrayrulecolor[gray]{0.5}\hhline{~|-------------}\arrayrulecolor{black}
& \baselinename{Oracle} 
& 52.60 & 62.64 & 66.16 & 60.24 & 70.49 & 73.87 
& 24.97 & 43.01 & 50.54 & 35.78 & 61.33 & 71.36  \\
\rowcolor{namerow}\cellcolor{white} & \ +\name                  
& 56.39 & 66.97 & 69.87 & 63.37 & 74.32 & 77.92 
& 38.18 & 53.64 & 59.73 & 58.93 & 77.22 & 82.37  \\
\midrule

\multirow{10}{*}{\shortstack{\fontfamily{pcr}\selectfont LLaMA\\\fontfamily{pcr}\selectfont -8B}}
& \baselinename{Vanilla} 
& 15.86 & 23.50 & 26.55 & 23.14 & 34.38 & 38.65 
&  8.13 & 16.97 & 21.73 & 22.78 & 44.36 & 55.25  \\
\rowcolor{namerow}\cellcolor{white} & \ +\name                  
& 32.31 & 45.28 & 50.57 & 50.18 & 65.69 & 70.92 
& 23.58 & 35.34 & 40.84 & 65.94 & 82.92 & 87.90  \\
[-0.3ex]\arrayrulecolor[gray]{0.5}\hhline{~|-------------}\arrayrulecolor{black}
& \baselinename{Naive RAG} 
& 15.74 & 23.71 & 26.74 & 30.77 & 42.88 & 46.70 
&  6.79 & 15.89 & 21.88 & 19.73 & 41.44 & 53.16  \\
\rowcolor{namerow}\cellcolor{white} & \ +\name                  
& 22.78 & 35.51 & 40.91 & 46.39 & 62.77 & 68.62 
& 25.13 & 37.72 & 43.92 & 65.56 & 82.12 & 87.14  \\
[-0.3ex]\arrayrulecolor[gray]{0.5}\hhline{~|-------------}\arrayrulecolor{black}
& \baselinename{EpiGen} 
& 11.72 & 19.86 & 22.93 & 23.55 & 38.04 & 43.94 
&  9.57 & 19.55 & 25.32 & 24.97 & 48.97 & 60.37  \\
& \ +\name                  
& \cellcolor{namerow} 27.51 & \cellcolor{namerow} 41.07 & \cellcolor{namerow} 46.99 & \cellcolor{namerow} 45.80 & \cellcolor{namerow} 61.75 & \cellcolor{namerow} 68.08 
& \cellcolor{namerow} 23.96 & \cellcolor{namerow} 36.33 & \cellcolor{namerow} 42.28 & \cellcolor{namerow} 64.49 & \cellcolor{namerow} 82.02 & \cellcolor{namerow} 87.04  \\
[-0.3ex]\arrayrulecolor[gray]{0.5}\hhline{~|-------------}\arrayrulecolor{black}
& \baselinename{CAPIR} 
& 13.20 & 19.97 & 22.69 & 22.25 & 34.77 & 40.18 
&  8.40 & 18.19 & 23.86 & 25.67 & 49.08 & 60.62  \\
\rowcolor{namerow}\cellcolor{white} & \ +\name                  
& 30.24 & 42.61 & 47.17 & 48.17 & 63.47 & 68.59 
& 23.64 & 35.16 & 40.36 & 66.20 & 82.72 & 87.42  \\
[-0.3ex]\arrayrulecolor[gray]{0.5}\hhline{~|-------------}\arrayrulecolor{black}
& \baselinename{Oracle} 
& 30.36 & 42.79 & 47.45 & 46.57 & 64.42 & 70.38 
& 13.10 & 26.73 & 33.66 & 29.36 & 52.52 & 62.55  \\
\rowcolor{namerow}\cellcolor{white} & \ +\name                  
& 36.75 & 51.93 & 57.64 & 55.98 & 75.35 & 81.29 
& 27.75 & 42.45 & 49.09 & 67.54 & 84.60 & 89.39  \\
\bottomrule
\end{tabular}
\end{adjustbox}

\end{table*}

\ding{182} \textit{\name improves private-library-oriented code generation.}
Across models and benchmarks, \name improves both \pass{k} and \exec{k} over the corresponding vanilla models in all reported settings.
For example, on \ndonnx with \qwen, \name increases \pass{1} from 23.37\% to 47.99\% and improves \exec{1} from 28.52\% to 55.86\%.
On \numba with \llama, \name increases \pass{5} from 21.73\% to 40.84\% and \exec{5} from 55.25\% to 87.90\%.
\seclrevision{\name alone also outperforms \baselinename{Oracle} on all metrics in five of the six model--benchmark settings, with \baselinename{Oracle} leading only for \qwen on \ndonnx.}
These improvements indicate that training on the synthesized data produced by \name helps models learn private-library API invocation patterns.
By comparison, the RAG-based baselines lead to smaller changes in many settings.
For instance, on \ndonnx with \qwen, \baselinename{EpiGen} improves \pass{1} and \exec{1} by 5.45 and 5.21 percentage points, respectively.

\ding{183} \textit{\name is complementary to RAG-based approaches.}
Combining \name with RAG-based approaches often yields higher metric values than using RAG alone.
For example, when combined with \baselinename{CAPIR}, \name achieves higher \pass{5} and \exec{5} on \ndonnx with \deepseek than either \name or \baselinename{CAPIR} alone.
Moreover, combining \name with \baselinename{Oracle} achieves the highest results in nearly all settings, suggesting that learning API invocation and providing task-specific API knowledge are complementary.
\lrevision{We also observe that combining \name with non-oracle RAG can underperform \name alone on many metrics.
For example, on \ndonnx with \deepseek, \name alone achieves 48.17\% \pass{1}, while combining \name with \baselinename{CAPIR} achieves 47.10\% \pass{1}.
We believe that two factors may account for this result.
First, imperfect retrieval may return irrelevant or low-quality API information that does not match the task. The model must then distinguish useful API specifications from this noisy context. When it fails to do so, the retrieved information can steer generation toward wrong APIs, offsetting the benefit of \name.
Table~\ref{tab:rag-required-api-recall} further shows that all three non-oracle RAG methods retrieve only a small fraction of the APIs required by each task, particularly on \numba.
Second, \name is trained without API context in Eq.~\ref{eq:sft} but receives retrieved API information at inference time in Eq.~\ref{eq:deploy_rag}, creating a train--inference mismatch that may limit its ability to exploit the additional context.
Improving retrieval quality and reducing this mismatch may therefore help the two approaches combine more effectively. We leave both directions to future work.}

\begin{table}[!tbp]
\crevision
\centering
\small
\renewcommand{\arraystretch}{1.02}
\setlength{\tabcolsep}{20pt}
\caption{\lrevision{Recall (\%) of the APIs required by each task for the three non-oracle RAG methods.}}
\label{tab:rag-required-api-recall}

\begin{adjustbox}{max width=\linewidth}
\begin{tabular}{llr}
\toprule
\textbf{Benchmark} & \textbf{Method} & \textbf{Recall} \\
\midrule
\multirow{3}{*}{\ndonnx}
& \baselinename{Naive RAG} & 22.94 \\
& \baselinename{CAPIR} & 18.76 \\
& \baselinename{EpiGen} & 19.41 \\
\midrule
\multirow{3}{*}{\numba}
& \baselinename{Naive RAG} & 3.87 \\
& \baselinename{CAPIR} & 10.36 \\
& \baselinename{EpiGen} & 4.80 \\
\bottomrule
\end{tabular}
\end{adjustbox}

\end{table}

\begin{table*}[!tbp]
\crevision
\centering
\small
\renewcommand{\arraystretch}{1.0}
\setlength{\tabcolsep}{8pt}
\caption{\lrevision{Direct measures of private-library API use on \ndonnx and \numba. \textbf{Inv.}, \textbf{Rec.}, \textbf{N.E.}, and \textbf{S.E.} denote the library invocation rate, required API recall, non-existent-call rate, and signature-error rate, respectively. All values are percentages.}}
\label{tab:rq1-api-use}

\begin{adjustbox}{max width=\linewidth}
\begin{tabular}{c|l|cccc|cccc}
\toprule
\multirow{2}{*}{\textbf{Model}} & \multirow{2}{*}{\textbf{Method}} &
\multicolumn{4}{c|}{\textbf{\ndonnx}} &
\multicolumn{4}{c}{\textbf{\numba}} \\
\cmidrule(lr){3-6}\cmidrule(lr){7-10}
& &
\textbf{Inv.}\,$\uparrow$ & \textbf{Rec.}\,$\uparrow$ & \textbf{N.E.}\,$\downarrow$ & \textbf{S.E.}\,$\downarrow$ &
\textbf{Inv.}\,$\uparrow$ & \textbf{Rec.}\,$\uparrow$ & \textbf{N.E.}\,$\downarrow$ & \textbf{S.E.}\,$\downarrow$ \\
\midrule

\multirow{3}{*}{\shortstack{\fontfamily{pcr}\selectfont DeepSeek\\\fontfamily{pcr}\selectfont-6.7B}}
& \baselinename{Vanilla}
& 92.96 & 20.25 & 22.73 & 2.78
& 92.25 & 21.98 & 6.76 & 1.02 \\
& \baselinename{Oracle}
& 86.98 & 48.85 & \textbf{0.48} & 2.60
& 93.64 & 45.55 & \textbf{2.20} & 3.42 \\
& \name
& \textbf{98.11} & \textbf{66.38} & 3.40 & \textbf{1.54}
& \textbf{95.24} & \textbf{49.64} & 4.08 & \textbf{0.91} \\
\midrule

\multirow{3}{*}{\shortstack{\fontfamily{pcr}\selectfont Qwen\\\fontfamily{pcr}\selectfont-7B}}
& \baselinename{Vanilla}
& 95.86 & 22.55 & 25.58 & 4.73
& 91.93 & 27.98 & 7.81 & 1.04 \\
& \baselinename{Oracle}
& 99.88 & 47.38 & \textbf{2.02} & \textbf{1.24}
& 90.21 & 54.29 & \textbf{2.87} & 2.62 \\
& \name
& \textbf{99.94} & \textbf{53.50} & 5.57 & 1.60
& \textbf{97.38} & \textbf{58.31} & 3.42 & \textbf{1.02} \\
\midrule

\multirow{3}{*}{\shortstack{\fontfamily{pcr}\selectfont LLaMA\\\fontfamily{pcr}\selectfont-8B}}
& \baselinename{Vanilla}
& 97.69 & 20.56 & 30.50 & 7.81
& 91.23 & 28.20 & 10.26 & 1.07 \\
& \baselinename{Oracle}
& 94.62 & 44.37 & \textbf{2.25} & 3.85
& 93.90 & 57.85 & \textbf{2.06} & 1.76 \\
& \name
& \textbf{100.00} & \textbf{66.58} & 4.00 & \textbf{2.25}
& \textbf{98.34} & \textbf{59.07} & 2.13 & \textbf{0.43} \\
\midrule
\multirow{3}{*}{\textbf{Average}}
& \baselinename{Vanilla}
& 95.50 & 21.12 & 26.27 & 5.11
& 91.80 & 26.05 & 8.28 & 1.04 \\
& \baselinename{Oracle}
& 93.83 & 46.87 & \textbf{1.58} & 2.56
& 92.58 & 52.56 & \textbf{2.38} & 2.60 \\
& \name
& \textbf{99.35} & \textbf{62.15} & 4.32 & \textbf{1.80}
& \textbf{96.99} & \textbf{55.67} & 3.21 & \textbf{0.79} \\
\bottomrule
\end{tabular}
\end{adjustbox}

\end{table*}

\begin{prevision}
\ding{184} \textit{The gains of \name reflect improved API invocation rather than only better standalone code generation.}
As shown in Section~\ref{sec:motivation}, LLMs frequently fail either to actively invoke private-library APIs or to invoke them correctly. To characterize these failures beyond \pass{k} and \exec{k}, we introduce four metrics. The library invocation rate measures whether a solution invokes the target library; required API recall measures coverage of the required API set; the non-existent-call rate measures the proportion of target-library calls to APIs absent from the library; and the signature-error rate measures whether APIs are invoked correctly by identifying calls that violate declared signatures. We compute these metrics for \baselinename{Vanilla}, \baselinename{Oracle}, and \name using the same generations evaluated in Table~\ref{tab:rq1}.

As shown in Table~\ref{tab:rq1-api-use}, \name improves all four average metrics over \baselinename{Vanilla} on both benchmarks. On \ndonnx, it increases the library invocation rate from 95.50\% to 99.35\% and required API recall from 21.12\% to 62.15\%, while reducing the non-existent-call rate from 26.27\% to 4.32\% and the signature-error rate from 5.11\% to 1.80\%. On \numba, the corresponding values improve from 91.80\% to 96.99\%, 26.05\% to 55.67\%, 8.28\% to 3.21\%, and 1.04\% to 0.79\%, respectively. Compared with \baselinename{Oracle}, \name achieves a higher library invocation rate and required API recall, and a lower signature-error rate on both benchmarks. Together, these results indicate that \name enables models to use private libraries more actively and correctly. However, \name has a higher non-existent-call rate than \baselinename{Oracle} on both benchmarks. This difference is expected because the \baselinename{Oracle} prompt explicitly states that the provided APIs are all the APIs required for the task, making the model less likely to invoke APIs beyond those provided. These results show that the gains in Table~\ref{tab:rq1} reflect improved private-library API invocation rather than only cleaner standalone Python code.
\end{prevision}

\begin{boxK}
\small \lrevision{\textbf{Answer to RQ1:}
Across both benchmarks, \name improves private-library API use over \baselinename{Vanilla}, with higher library invocation rates and required API recall and fewer non-existent calls and signature errors. It is also complementary to RAG-based approaches.}
\end{boxK}

\subsection{RQ2: Does \name negatively affect general code generation capability?}
\label{sec:rq2}

\begin{table}[!tbp]

\centering
\small
\renewcommand{\arraystretch}{1.0}
\setlength{\tabcolsep}{2.0pt}
\caption{\pass{k} and \exec{k} (\%) on \humaneval under two private-library settings.
For retrieval-based baselines, the retrieved corpus is the API documentation of the corresponding library; \name is trained on synthesized data constructed for the same library.}
\label{tab:rq2}

\begin{subtable}[t]{0.492\textwidth}
\centering
\caption{\texttt{numba-cuda} setting}
\label{tab:rq2-numba}
\resizebox{\linewidth}{!}{
\begin{tabular}{@{}lcccc@{}}
\toprule
\textbf{Method} & \textbf{\pass{1}} & \textbf{\pass{5}} & \textbf{\exec{1}} & \textbf{\exec{5}} \\
\midrule
\multicolumn{5}{c}{\textbf{\deepseek}} \\
\arrayrulecolor{gray}\midrule\arrayrulecolor{black}
\rowcolor{vanillarow}
\baselinename{Vanilla}
& 71.89 \textcolor[HTML]{909090}{\scriptsize (-0.00)}
& 86.96 \textcolor[HTML]{909090}{\scriptsize (-0.00)}
& 93.60 \textcolor[HTML]{909090}{\scriptsize (-0.00)}
& 98.27 \textcolor[HTML]{909090}{\scriptsize (-0.00)} \\
\rowcolor{ragrow}
\baselinename{Naive RAG}
& 47.13 \textcolor[HTML]{0B3D91}{\scriptsize (-24.76)}
& 70.88 \textcolor[HTML]{0B3D91}{\scriptsize (-16.08)}
& 64.45 \textcolor[HTML]{0B3D91}{\scriptsize (-29.15)}
& 88.65 \textcolor[HTML]{0B3D91}{\scriptsize (-9.62)} \\
\rowcolor{ragrow}
\baselinename{EpiGen}
& 48.41 \textcolor[HTML]{0B3D91}{\scriptsize (-23.48)}
& 72.88 \textcolor[HTML]{0B3D91}{\scriptsize (-14.08)}
& 64.82 \textcolor[HTML]{0B3D91}{\scriptsize (-28.78)}
& 89.54 \textcolor[HTML]{0B3D91}{\scriptsize (-8.73)} \\
\rowcolor{ragrow}
\baselinename{CAPIR}
& 46.52 \textcolor[HTML]{0B3D91}{\scriptsize (-25.37)}
& 71.29 \textcolor[HTML]{0B3D91}{\scriptsize (-15.67)}
& 67.20 \textcolor[HTML]{0B3D91}{\scriptsize (-26.40)}
& 86.90 \textcolor[HTML]{0B3D91}{\scriptsize (-11.37)} \\
\rowcolor{newnamerow}
\baselinename{\name}
& 69.82 \textcolor[HTML]{0B3D91}{\scriptsize (-2.07)}
& 85.92 \textcolor[HTML]{0B3D91}{\scriptsize (-1.04)}
& 95.30 \textcolor[HTML]{8B0000}{\scriptsize (+1.70)}
& 98.37 \textcolor[HTML]{8B0000}{\scriptsize (+0.10)} \\
\midrule
\multicolumn{5}{c}{\textbf{\qwen}} \\
\arrayrulecolor{gray}\midrule\arrayrulecolor{black}
\rowcolor{vanillarow}
\baselinename{Vanilla}
& 85.43 \textcolor[HTML]{909090}{\scriptsize (-0.00)}
& 92.85 \textcolor[HTML]{909090}{\scriptsize (-0.00)}
& 98.29 \textcolor[HTML]{909090}{\scriptsize (-0.00)}
& 99.95 \textcolor[HTML]{909090}{\scriptsize (-0.00)} \\
\rowcolor{ragrow}
\baselinename{Naive RAG}
& 58.96 \textcolor[HTML]{0B3D91}{\scriptsize (-26.47)}
& 77.91 \textcolor[HTML]{0B3D91}{\scriptsize (-14.94)}
& 70.79 \textcolor[HTML]{0B3D91}{\scriptsize (-27.50)}
& 88.91 \textcolor[HTML]{0B3D91}{\scriptsize (-11.04)} \\
\rowcolor{ragrow}
\baselinename{EpiGen}
& 58.23 \textcolor[HTML]{0B3D91}{\scriptsize (-27.20)}
& 77.23 \textcolor[HTML]{0B3D91}{\scriptsize (-15.62)}
& 69.51 \textcolor[HTML]{0B3D91}{\scriptsize (-28.78)}
& 87.14 \textcolor[HTML]{0B3D91}{\scriptsize (-12.81)} \\
\rowcolor{ragrow}
\baselinename{CAPIR}
& 53.17 \textcolor[HTML]{0B3D91}{\scriptsize (-32.26)}
& 72.17 \textcolor[HTML]{0B3D91}{\scriptsize (-20.68)}
& 65.91 \textcolor[HTML]{0B3D91}{\scriptsize (-32.38)}
& 84.33 \textcolor[HTML]{0B3D91}{\scriptsize (-15.62)} \\
\rowcolor{newnamerow}
\baselinename{\name}
& 84.02 \textcolor[HTML]{0B3D91}{\scriptsize (-1.41)}
& 90.99 \textcolor[HTML]{0B3D91}{\scriptsize (-1.86)}
& 98.23 \textcolor[HTML]{0B3D91}{\scriptsize (-0.06)}
& 99.78 \textcolor[HTML]{0B3D91}{\scriptsize (-0.17)} \\
\midrule
\multicolumn{5}{c}{\textbf{\llama}} \\
\arrayrulecolor{gray}\midrule\arrayrulecolor{black}
\rowcolor{vanillarow}
\baselinename{Vanilla}
& 64.82 \textcolor[HTML]{909090}{\scriptsize (-0.00)}
& 78.17 \textcolor[HTML]{909090}{\scriptsize (-0.00)}
& 95.37 \textcolor[HTML]{909090}{\scriptsize (-0.00)}
& 99.29 \textcolor[HTML]{909090}{\scriptsize (-0.00)} \\
\rowcolor{ragrow}
\baselinename{Naive RAG}
& 46.59 \textcolor[HTML]{0B3D91}{\scriptsize (-18.23)}
& 72.13 \textcolor[HTML]{0B3D91}{\scriptsize (-6.04)}
& 75.67 \textcolor[HTML]{0B3D91}{\scriptsize (-19.70)}
& 94.50 \textcolor[HTML]{0B3D91}{\scriptsize (-4.79)} \\
\rowcolor{ragrow}
\baselinename{EpiGen}
& 46.40 \textcolor[HTML]{0B3D91}{\scriptsize (-18.42)}
& 70.78 \textcolor[HTML]{0B3D91}{\scriptsize (-7.39)}
& 78.84 \textcolor[HTML]{0B3D91}{\scriptsize (-16.53)}
& 95.01 \textcolor[HTML]{0B3D91}{\scriptsize (-4.28)} \\
\rowcolor{ragrow}
\baselinename{CAPIR}
& 39.88 \textcolor[HTML]{0B3D91}{\scriptsize (-24.94)}
& 63.93 \textcolor[HTML]{0B3D91}{\scriptsize (-14.24)}
& 66.34 \textcolor[HTML]{0B3D91}{\scriptsize (-29.03)}
& 87.56 \textcolor[HTML]{0B3D91}{\scriptsize (-11.73)} \\
\rowcolor{newnamerow}
\baselinename{\name}
& 65.12 \textcolor[HTML]{8B0000}{\scriptsize (+0.30)}
& 78.94 \textcolor[HTML]{8B0000}{\scriptsize (+0.77)}
& 96.46 \textcolor[HTML]{8B0000}{\scriptsize (+1.09)}
& 99.56 \textcolor[HTML]{8B0000}{\scriptsize (+0.27)} \\
\bottomrule
\end{tabular}
}
\end{subtable}
\hfill
\begin{subtable}[t]{0.492\textwidth}
\centering
\caption{\texttt{ndonnx} setting}
\label{tab:rq2-ndonnx}
\resizebox{\linewidth}{!}{
\begin{tabular}{@{}lcccc@{}}
\toprule
\textbf{Method} & \textbf{\pass{1}} & \textbf{\pass{5}} & \textbf{\exec{1}} & \textbf{\exec{5}} \\
\midrule
\multicolumn{5}{c}{\textbf{\deepseek}} \\
\arrayrulecolor{gray}\midrule\arrayrulecolor{black}
\rowcolor{vanillarow}
\baselinename{Vanilla}
& 71.89 \textcolor[HTML]{909090}{\scriptsize (-0.00)}
& 86.96 \textcolor[HTML]{909090}{\scriptsize (-0.00)}
& 93.60 \textcolor[HTML]{909090}{\scriptsize (-0.00)}
& 98.27 \textcolor[HTML]{909090}{\scriptsize (-0.00)} \\
\rowcolor{ragrow}
\baselinename{Naive RAG}
& 50.00 \textcolor[HTML]{0B3D91}{\scriptsize (-21.89)}
& 76.99 \textcolor[HTML]{0B3D91}{\scriptsize (-9.97)}
& 71.52 \textcolor[HTML]{0B3D91}{\scriptsize (-22.08)}
& 92.24 \textcolor[HTML]{0B3D91}{\scriptsize (-6.03)} \\
\rowcolor{ragrow}
\baselinename{EpiGen}
& 51.22 \textcolor[HTML]{0B3D91}{\scriptsize (-20.67)}
& 77.55 \textcolor[HTML]{0B3D91}{\scriptsize (-9.41)}
& 73.41 \textcolor[HTML]{0B3D91}{\scriptsize (-20.19)}
& 94.55 \textcolor[HTML]{0B3D91}{\scriptsize (-3.72)} \\
\rowcolor{ragrow}
\baselinename{CAPIR}
& 56.89 \textcolor[HTML]{0B3D91}{\scriptsize (-15.00)}
& 81.35 \textcolor[HTML]{0B3D91}{\scriptsize (-5.61)}
& 80.73 \textcolor[HTML]{0B3D91}{\scriptsize (-12.87)}
& 96.24 \textcolor[HTML]{0B3D91}{\scriptsize (-2.03)} \\
\rowcolor{newnamerow}
\baselinename{\name}
& 65.79 \textcolor[HTML]{0B3D91}{\scriptsize (-6.10)}
& 83.55 \textcolor[HTML]{0B3D91}{\scriptsize (-3.41)}
& 91.04 \textcolor[HTML]{0B3D91}{\scriptsize (-2.56)}
& 97.08 \textcolor[HTML]{0B3D91}{\scriptsize (-1.19)} \\
\midrule
\multicolumn{5}{c}{\textbf{\qwen}} \\
\arrayrulecolor{gray}\midrule\arrayrulecolor{black}
\rowcolor{vanillarow}
\baselinename{Vanilla}
& 85.43 \textcolor[HTML]{909090}{\scriptsize (-0.00)}
& 92.85 \textcolor[HTML]{909090}{\scriptsize (-0.00)}
& 98.29 \textcolor[HTML]{909090}{\scriptsize (-0.00)}
& 99.95 \textcolor[HTML]{909090}{\scriptsize (-0.00)} \\
\rowcolor{ragrow}
\baselinename{Naive RAG}
& 69.21 \textcolor[HTML]{0B3D91}{\scriptsize (-16.22)}
& 82.85 \textcolor[HTML]{0B3D91}{\scriptsize (-10.00)}
& 78.66 \textcolor[HTML]{0B3D91}{\scriptsize (-19.63)}
& 89.47 \textcolor[HTML]{0B3D91}{\scriptsize (-10.48)} \\
\rowcolor{ragrow}
\baselinename{EpiGen}
& 65.06 \textcolor[HTML]{0B3D91}{\scriptsize (-20.37)}
& 81.29 \textcolor[HTML]{0B3D91}{\scriptsize (-11.56)}
& 74.94 \textcolor[HTML]{0B3D91}{\scriptsize (-23.35)}
& 86.98 \textcolor[HTML]{0B3D91}{\scriptsize (-12.97)} \\
\rowcolor{ragrow}
\baselinename{CAPIR}
& 68.48 \textcolor[HTML]{0B3D91}{\scriptsize (-16.95)}
& 87.41 \textcolor[HTML]{0B3D91}{\scriptsize (-5.44)}
& 78.90 \textcolor[HTML]{0B3D91}{\scriptsize (-19.39)}
& 92.62 \textcolor[HTML]{0B3D91}{\scriptsize (-7.33)} \\
\rowcolor{newnamerow}
\baselinename{\name}
& 85.79 \textcolor[HTML]{8B0000}{\scriptsize (+0.36)}
& 92.01 \textcolor[HTML]{0B3D91}{\scriptsize (-0.84)}
& 97.80 \textcolor[HTML]{0B3D91}{\scriptsize (-0.49)}
& 99.90 \textcolor[HTML]{0B3D91}{\scriptsize (-0.05)} \\
\midrule
\multicolumn{5}{c}{\textbf{\llama}} \\
\arrayrulecolor{gray}\midrule\arrayrulecolor{black}
\rowcolor{vanillarow}
\baselinename{Vanilla}
& 64.82 \textcolor[HTML]{909090}{\scriptsize (-0.00)}
& 78.17 \textcolor[HTML]{909090}{\scriptsize (-0.00)}
& 95.37 \textcolor[HTML]{909090}{\scriptsize (-0.00)}
& 99.29 \textcolor[HTML]{909090}{\scriptsize (-0.00)} \\
\rowcolor{ragrow}
\baselinename{Naive RAG}
& 54.45 \textcolor[HTML]{0B3D91}{\scriptsize (-10.37)}
& 76.16 \textcolor[HTML]{0B3D91}{\scriptsize (-2.01)}
& 86.28 \textcolor[HTML]{0B3D91}{\scriptsize (-9.09)}
& 97.67 \textcolor[HTML]{0B3D91}{\scriptsize (-1.62)} \\
\rowcolor{ragrow}
\baselinename{EpiGen}
& 55.00 \textcolor[HTML]{0B3D91}{\scriptsize (-9.82)}
& 76.17 \textcolor[HTML]{0B3D91}{\scriptsize (-2.00)}
& 85.61 \textcolor[HTML]{0B3D91}{\scriptsize (-9.76)}
& 97.74 \textcolor[HTML]{0B3D91}{\scriptsize (-1.55)} \\
\rowcolor{ragrow}
\baselinename{CAPIR}
& 53.96 \textcolor[HTML]{0B3D91}{\scriptsize (-10.86)}
& 75.47 \textcolor[HTML]{0B3D91}{\scriptsize (-2.70)}
& 84.57 \textcolor[HTML]{0B3D91}{\scriptsize (-10.80)}
& 97.67 \textcolor[HTML]{0B3D91}{\scriptsize (-1.62)} \\
\rowcolor{newnamerow}
\baselinename{\name}
& 62.44 \textcolor[HTML]{0B3D91}{\scriptsize (-2.38)}
& 76.01 \textcolor[HTML]{0B3D91}{\scriptsize (-2.16)}
& 93.11 \textcolor[HTML]{0B3D91}{\scriptsize (-2.26)}
& 98.65 \textcolor[HTML]{0B3D91}{\scriptsize (-0.64)} \\
\bottomrule
\end{tabular}
}
\end{subtable}

\end{table}

Although \name substantially improves models' ability to use private libraries, real-world development involves many requirements that do not rely on private-library APIs. In RQ2, we evaluate whether training with \name compromises models' general code generation capability.

\vspace{2pt}
\noindent \textbf{Setting.}
We apply the baselines and \name to the three models described in Section~\ref{sec:model} and evaluate them on \humaneval \lrevision{and \mbpp}.
Although \humaneval \lrevision{and \mbpp do} not require private-library APIs, we still enable each private-library-oriented method in this evaluation to simulate realistic development workflows: developers and deployed coding assistants often do not know in advance whether a given requirement should use private libraries.
Thus, this RQ serves as a stress test of whether always-available private-library support interferes with ordinary code generation.
For RAG-based baselines, the retriever searches the documentation of the corresponding private library for each \lrevision{requirement from both benchmarks}. \lrevision{For} \name, we directly use the model fine-tuned on the corresponding synthesized private-library data.
We report \pass{k} and \exec{k} as evaluation metrics, where $k\in\{1,5\}$.

\vspace{2pt}
\noindent \textbf{Results.}
\lrevision{Tables}~\ref{tab:rq2} \lrevision{and~\ref{tab:rq2-mbpp}} report \pass{k} and \exec{k} on \humaneval \lrevision{and \mbpp} under both the \texttt{numba-cuda} and \texttt{ndonnx} settings.

\begin{table}[!tbp]
\crevision

\centering
\small
\renewcommand{\arraystretch}{1.0}
\setlength{\tabcolsep}{2.0pt}
\caption{
\lrevision{\pass{k} and \exec{k} (\%) on \mbpp under two private-library settings.
For retrieval-based baselines, the retrieved corpus is the API documentation of the corresponding library. \name is trained on synthesized data constructed for the same library.}
}
\label{tab:rq2-mbpp}

\begin{subtable}[t]{0.492\textwidth}
\centering
\caption{\texttt{numba-cuda} setting}
\label{tab:rq2-mbpp-numba}
\resizebox{\linewidth}{!}{
\begin{tabular}{@{}lcccc@{}}
\toprule
\textbf{Method} & \textbf{\pass{1}} & \textbf{\pass{5}} & \textbf{\exec{1}} & \textbf{\exec{5}} \\
\midrule
\multicolumn{5}{c}{\textbf{\deepseek}} \\
\arrayrulecolor{gray}\midrule\arrayrulecolor{black}
\rowcolor{vanillarow}
\baselinename{Vanilla}
& 66.46 \textcolor[HTML]{909090}{\scriptsize (-0.00)}
& 78.79 \textcolor[HTML]{909090}{\scriptsize (-0.00)}
& 89.79 \textcolor[HTML]{909090}{\scriptsize (-0.00)}
& 95.82 \textcolor[HTML]{909090}{\scriptsize (-0.00)} \\
\rowcolor{ragrow}
\baselinename{Naive RAG}
& 42.78 \textcolor[HTML]{0B3D91}{\scriptsize (-23.68)}
& 62.81 \textcolor[HTML]{0B3D91}{\scriptsize (-15.98)}
& 57.43 \textcolor[HTML]{0B3D91}{\scriptsize (-32.36)}
& 79.18 \textcolor[HTML]{0B3D91}{\scriptsize (-16.64)} \\
\rowcolor{ragrow}
\baselinename{EpiGen}
& 44.13 \textcolor[HTML]{0B3D91}{\scriptsize (-22.33)}
& 64.76 \textcolor[HTML]{0B3D91}{\scriptsize (-14.03)}
& 58.47 \textcolor[HTML]{0B3D91}{\scriptsize (-31.32)}
& 79.82 \textcolor[HTML]{0B3D91}{\scriptsize (-16.00)} \\
\rowcolor{ragrow}
\baselinename{CAPIR}
& 43.52 \textcolor[HTML]{0B3D91}{\scriptsize (-22.94)}
& 63.13 \textcolor[HTML]{0B3D91}{\scriptsize (-15.66)}
& 58.57 \textcolor[HTML]{0B3D91}{\scriptsize (-31.22)}
& 78.36 \textcolor[HTML]{0B3D91}{\scriptsize (-17.46)} \\
\rowcolor{newnamerow}
\baselinename{\name}
& 60.08 \textcolor[HTML]{0B3D91}{\scriptsize (-6.38)}
& 78.76 \textcolor[HTML]{0B3D91}{\scriptsize (-0.03)}
& 89.71 \textcolor[HTML]{0B3D91}{\scriptsize (-0.08)}
& 96.71 \textcolor[HTML]{8B0000}{\scriptsize (+0.89)} \\
\midrule
\multicolumn{5}{c}{\textbf{\qwen}} \\
\arrayrulecolor{gray}\midrule\arrayrulecolor{black}
\rowcolor{vanillarow}
\baselinename{Vanilla}
& 82.12 \textcolor[HTML]{909090}{\scriptsize (-0.00)}
& 92.06 \textcolor[HTML]{909090}{\scriptsize (-0.00)}
& 98.62 \textcolor[HTML]{909090}{\scriptsize (-0.00)}
& 99.94 \textcolor[HTML]{909090}{\scriptsize (-0.00)} \\
\rowcolor{ragrow}
\baselinename{Naive RAG}
& 50.69 \textcolor[HTML]{0B3D91}{\scriptsize (-31.43)}
& 76.18 \textcolor[HTML]{0B3D91}{\scriptsize (-15.88)}
& 63.28 \textcolor[HTML]{0B3D91}{\scriptsize (-35.34)}
& 84.83 \textcolor[HTML]{0B3D91}{\scriptsize (-15.11)} \\
\rowcolor{ragrow}
\baselinename{EpiGen}
& 56.48 \textcolor[HTML]{0B3D91}{\scriptsize (-25.64)}
& 77.71 \textcolor[HTML]{0B3D91}{\scriptsize (-14.35)}
& 68.81 \textcolor[HTML]{0B3D91}{\scriptsize (-29.81)}
& 85.97 \textcolor[HTML]{0B3D91}{\scriptsize (-13.97)} \\
\rowcolor{ragrow}
\baselinename{CAPIR}
& 69.23 \textcolor[HTML]{0B3D91}{\scriptsize (-12.89)}
& 84.29 \textcolor[HTML]{0B3D91}{\scriptsize (-7.77)}
& 83.20 \textcolor[HTML]{0B3D91}{\scriptsize (-15.42)}
& 92.88 \textcolor[HTML]{0B3D91}{\scriptsize (-7.06)} \\
\rowcolor{newnamerow}
\baselinename{\name}
& 80.13 \textcolor[HTML]{0B3D91}{\scriptsize (-1.99)}
& 90.69 \textcolor[HTML]{0B3D91}{\scriptsize (-1.37)}
& 99.05 \textcolor[HTML]{8B0000}{\scriptsize (+0.43)}
& 99.71 \textcolor[HTML]{0B3D91}{\scriptsize (-0.23)} \\
\midrule
\multicolumn{5}{c}{\textbf{\llama}} \\
\arrayrulecolor{gray}\midrule\arrayrulecolor{black}
\rowcolor{vanillarow}
\baselinename{Vanilla}
& 69.97 \textcolor[HTML]{909090}{\scriptsize (-0.00)}
& 79.58 \textcolor[HTML]{909090}{\scriptsize (-0.00)}
& 95.71 \textcolor[HTML]{909090}{\scriptsize (-0.00)}
& 97.70 \textcolor[HTML]{909090}{\scriptsize (-0.00)} \\
\rowcolor{ragrow}
\baselinename{Naive RAG}
& 57.46 \textcolor[HTML]{0B3D91}{\scriptsize (-12.51)}
& 77.27 \textcolor[HTML]{0B3D91}{\scriptsize (-2.31)}
& 85.16 \textcolor[HTML]{0B3D91}{\scriptsize (-10.55)}
& 96.33 \textcolor[HTML]{0B3D91}{\scriptsize (-1.37)} \\
\rowcolor{ragrow}
\baselinename{EpiGen}
& 55.19 \textcolor[HTML]{0B3D91}{\scriptsize (-14.78)}
& 76.43 \textcolor[HTML]{0B3D91}{\scriptsize (-3.15)}
& 82.51 \textcolor[HTML]{0B3D91}{\scriptsize (-13.20)}
& 95.57 \textcolor[HTML]{0B3D91}{\scriptsize (-2.13)} \\
\rowcolor{ragrow}
\baselinename{CAPIR}
& 51.38 \textcolor[HTML]{0B3D91}{\scriptsize (-18.59)}
& 74.48 \textcolor[HTML]{0B3D91}{\scriptsize (-5.10)}
& 77.06 \textcolor[HTML]{0B3D91}{\scriptsize (-18.65)}
& 93.74 \textcolor[HTML]{0B3D91}{\scriptsize (-3.96)} \\
\rowcolor{newnamerow}
\baselinename{\name}
& 69.18 \textcolor[HTML]{0B3D91}{\scriptsize (-0.79)}
& 80.87 \textcolor[HTML]{8B0000}{\scriptsize (+1.29)}
& 93.65 \textcolor[HTML]{0B3D91}{\scriptsize (-2.06)}
& 96.90 \textcolor[HTML]{0B3D91}{\scriptsize (-0.80)} \\
\bottomrule
\end{tabular}
}
\end{subtable}
\hfill
\begin{subtable}[t]{0.492\textwidth}
\centering
\caption{\texttt{ndonnx} setting}
\label{tab:rq2-mbpp-ndonnx}
\resizebox{\linewidth}{!}{
\begin{tabular}{@{}lcccc@{}}
\toprule
\textbf{Method} & \textbf{\pass{1}} & \textbf{\pass{5}} & \textbf{\exec{1}} & \textbf{\exec{5}} \\
\midrule
\multicolumn{5}{c}{\textbf{\deepseek}} \\
\arrayrulecolor{gray}\midrule\arrayrulecolor{black}
\rowcolor{vanillarow}
\baselinename{Vanilla}
& 66.46 \textcolor[HTML]{909090}{\scriptsize (-0.00)}
& 78.79 \textcolor[HTML]{909090}{\scriptsize (-0.00)}
& 89.79 \textcolor[HTML]{909090}{\scriptsize (-0.00)}
& 95.82 \textcolor[HTML]{909090}{\scriptsize (-0.00)} \\
\rowcolor{ragrow}
\baselinename{Naive RAG}
& 47.62 \textcolor[HTML]{0B3D91}{\scriptsize (-18.84)}
& 70.56 \textcolor[HTML]{0B3D91}{\scriptsize (-8.23)}
& 65.13 \textcolor[HTML]{0B3D91}{\scriptsize (-24.66)}
& 87.34 \textcolor[HTML]{0B3D91}{\scriptsize (-8.48)} \\
\rowcolor{ragrow}
\baselinename{EpiGen}
& 47.59 \textcolor[HTML]{0B3D91}{\scriptsize (-18.87)}
& 69.92 \textcolor[HTML]{0B3D91}{\scriptsize (-8.87)}
& 67.14 \textcolor[HTML]{0B3D91}{\scriptsize (-22.65)}
& 88.51 \textcolor[HTML]{0B3D91}{\scriptsize (-7.31)} \\
\rowcolor{ragrow}
\baselinename{CAPIR}
& 46.08 \textcolor[HTML]{0B3D91}{\scriptsize (-20.38)}
& 65.90 \textcolor[HTML]{0B3D91}{\scriptsize (-12.89)}
& 64.87 \textcolor[HTML]{0B3D91}{\scriptsize (-24.92)}
& 85.07 \textcolor[HTML]{0B3D91}{\scriptsize (-10.75)} \\
\rowcolor{newnamerow}
\baselinename{\name}
& 64.37 \textcolor[HTML]{0B3D91}{\scriptsize (-2.09)}
& 77.28 \textcolor[HTML]{0B3D91}{\scriptsize (-1.51)}
& 87.49 \textcolor[HTML]{0B3D91}{\scriptsize (-2.30)}
& 93.96 \textcolor[HTML]{0B3D91}{\scriptsize (-1.86)} \\
\midrule
\multicolumn{5}{c}{\textbf{\qwen}} \\
\arrayrulecolor{gray}\midrule\arrayrulecolor{black}
\rowcolor{vanillarow}
\baselinename{Vanilla}
& 82.12 \textcolor[HTML]{909090}{\scriptsize (-0.00)}
& 92.06 \textcolor[HTML]{909090}{\scriptsize (-0.00)}
& 98.62 \textcolor[HTML]{909090}{\scriptsize (-0.00)}
& 99.94 \textcolor[HTML]{909090}{\scriptsize (-0.00)} \\
\rowcolor{ragrow}
\baselinename{Naive RAG}
& 68.99 \textcolor[HTML]{0B3D91}{\scriptsize (-13.13)}
& 86.74 \textcolor[HTML]{0B3D91}{\scriptsize (-5.32)}
& 83.41 \textcolor[HTML]{0B3D91}{\scriptsize (-15.21)}
& 95.29 \textcolor[HTML]{0B3D91}{\scriptsize (-4.65)} \\
\rowcolor{ragrow}
\baselinename{EpiGen}
& 69.02 \textcolor[HTML]{0B3D91}{\scriptsize (-13.10)}
& 86.68 \textcolor[HTML]{0B3D91}{\scriptsize (-5.38)}
& 83.28 \textcolor[HTML]{0B3D91}{\scriptsize (-15.34)}
& 94.59 \textcolor[HTML]{0B3D91}{\scriptsize (-5.35)} \\
\rowcolor{ragrow}
\baselinename{CAPIR}
& 73.65 \textcolor[HTML]{0B3D91}{\scriptsize (-8.47)}
& 88.61 \textcolor[HTML]{0B3D91}{\scriptsize (-3.45)}
& 87.72 \textcolor[HTML]{0B3D91}{\scriptsize (-10.90)}
& 96.00 \textcolor[HTML]{0B3D91}{\scriptsize (-3.94)} \\
\rowcolor{newnamerow}
\baselinename{\name}
& 79.87 \textcolor[HTML]{0B3D91}{\scriptsize (-2.25)}
& 90.38 \textcolor[HTML]{0B3D91}{\scriptsize (-1.68)}
& 97.38 \textcolor[HTML]{0B3D91}{\scriptsize (-1.24)}
& 99.46 \textcolor[HTML]{0B3D91}{\scriptsize (-0.48)} \\
\midrule
\multicolumn{5}{c}{\textbf{\llama}} \\
\arrayrulecolor{gray}\midrule\arrayrulecolor{black}
\rowcolor{vanillarow}
\baselinename{Vanilla}
& 69.97 \textcolor[HTML]{909090}{\scriptsize (-0.00)}
& 79.58 \textcolor[HTML]{909090}{\scriptsize (-0.00)}
& 95.71 \textcolor[HTML]{909090}{\scriptsize (-0.00)}
& 97.70 \textcolor[HTML]{909090}{\scriptsize (-0.00)} \\
\rowcolor{ragrow}
\baselinename{Naive RAG}
& 57.12 \textcolor[HTML]{0B3D91}{\scriptsize (-12.85)}
& 77.91 \textcolor[HTML]{0B3D91}{\scriptsize (-1.67)}
& 87.09 \textcolor[HTML]{0B3D91}{\scriptsize (-8.62)}
& 96.93 \textcolor[HTML]{0B3D91}{\scriptsize (-0.77)} \\
\rowcolor{ragrow}
\baselinename{EpiGen}
& 55.98 \textcolor[HTML]{0B3D91}{\scriptsize (-13.99)}
& 76.88 \textcolor[HTML]{0B3D91}{\scriptsize (-2.70)}
& 85.21 \textcolor[HTML]{0B3D91}{\scriptsize (-10.50)}
& 97.07 \textcolor[HTML]{0B3D91}{\scriptsize (-0.63)} \\
\rowcolor{ragrow}
\baselinename{CAPIR}
& 55.69 \textcolor[HTML]{0B3D91}{\scriptsize (-14.28)}
& 75.76 \textcolor[HTML]{0B3D91}{\scriptsize (-3.82)}
& 84.26 \textcolor[HTML]{0B3D91}{\scriptsize (-11.45)}
& 95.79 \textcolor[HTML]{0B3D91}{\scriptsize (-1.91)} \\
\rowcolor{newnamerow}
\baselinename{\name}
& 67.33 \textcolor[HTML]{0B3D91}{\scriptsize (-2.64)}
& 78.86 \textcolor[HTML]{0B3D91}{\scriptsize (-0.72)}
& 93.31 \textcolor[HTML]{0B3D91}{\scriptsize (-2.40)}
& 96.81 \textcolor[HTML]{0B3D91}{\scriptsize (-0.89)} \\
\bottomrule
\end{tabular}
}
\end{subtable}

\end{table}

\begin{figure}[!tbp]
    \centering
    \includegraphics[width=0.7\textwidth]{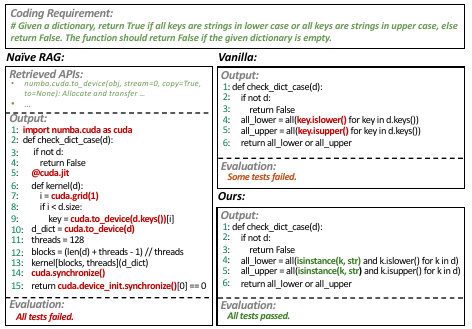}

    \caption{Case study on \humaneval with \llama under the \texttt{numba-cuda} setting.}
    \label{fig:rq2-cases}

\end{figure}

\ding{182} \textit{\name \lrevision{generally outperforms} RAG-based baselines on general code generation.}
\lrevision{On \humaneval under} the \texttt{numba-cuda} setting, for example, \name reduces the \pass{5} of \qwen by only 1.86 percentage points, whereas \baselinename{CAPIR} decreases it by 20.68 percentage points.
\lrevision{The \mbpp results show a similar advantage. With \qwen, \name achieves 80.13\% \pass{1} under the \texttt{numba-cuda} setting and 79.87\% under the \texttt{ndonnx} setting, compared with 69.23\% and 73.65\% for \baselinename{CAPIR}, respectively.}
We attribute the larger degradation of RAG-based baselines to the fact that they inject private-library knowledge into the context without determining whether such knowledge is actually needed.
This may encourage unnecessary private-library API usage even for tasks that should be solved with ordinary Python code.
Figure~\ref{fig:rq2-cases} illustrates such a failure case, where \baselinename{Naive RAG} incorrectly invokes \texttt{numba-cuda} APIs for a requirement that does not need them, leading to an incorrect solution.

\ding{183} \textit{\name can even enhance a model's general code generation capabilities in some cases.}
\lrevision{On \humaneval under} the \texttt{numba-cuda} setting, \name increases the \pass{5} of \llama from 78.17\% to 78.94\% and its \exec{1} from 95.37\% to 96.46\%.
\lrevision{On \mbpp, it also raises \llama's \pass{5} by 1.29 percentage points under the same setting.}
We attribute this unexpected benefit to the Multidimensional Graph Pruning. By exclusively training on data that has passed Multidimensional Graph Pruning, the model may reinforce its adherence to correct syntax and programming paradigms. As shown in Figure~\ref{fig:rq2-cases}, the \baselinename{Vanilla} model fails to use \colorbox{gray!15}{\texttt{isinstance}}, whereas the model trained with \name applies it correctly and successfully passes the test cases.

\begin{boxK}
\small \textbf{Answer to RQ2:}
\lrevision{Across \humaneval and \mbpp, \name generally outperforms RAG-based baselines and sometimes improves on the \baselinename{Vanilla} models.}
\end{boxK}

\subsection{RQ3: What is the contribution of each component in \name?}
\label{sec:rq3}

\name consists of two key components: Progressive Graph Evolution and Multidimensional Graph Pruning. In this RQ, we conduct an ablation study to examine how each component contributes to the overall effectiveness of \name.

\vspace{2pt}
\noindent \textbf{Setting.}
We compare four variants of \name.
\ding{182} \bi{Full \name} keeps both Progressive Graph Evolution and Multidimensional Graph Pruning.
\ding{183} \bi{w/o Graph Pruning} keeps Progressive Graph Evolution but removes Multidimensional Graph Pruning. As a result, all evolved sample nodes are directly used for training without filtering.
\ding{184} \bi{w/o Graph Evolution} keeps Multidimensional Graph Pruning but removes Progressive Graph Evolution. \lrevision{In this setting, we directly provide the LLM with the full API documentation and prompt it to synthesize training samples, without applying either Graph Initial Evolution or Graph Iterative Evolution.}
\ding{185} \bi{Direct Synthesis} removes both components and directly prompts the LLM with complete API documents to synthesize training samples for fine-tuning.

We conduct all ablation experiments on \qwen \lrevision{using \ndonnx, \numba, and \humaneval}.
For \ndonnx and \numba, we evaluate models trained on synthesized data constructed for \texttt{ndonnx} and \texttt{numba-cuda}, respectively.
For \humaneval, we evaluate the model trained on synthesized data constructed for \texttt{numba-cuda}.
We report \pass{k} and \exec{k}, where $k\in\{1,5\}$.
To ensure a fair comparison, all ablation variants use the same amount of training data.
We also include \baselinename{Vanilla} as a reference point to show the gains brought by different training strategies.

\begin{table*}[!tbp]
\centering
\small
\renewcommand{\arraystretch}{1.0}
\setlength{\tabcolsep}{1.15pt}
\caption{Ablation study of \name on \ndonnx and \numba. }
\label{tab:ablation}

\begin{adjustbox}{max width=\linewidth}
\begin{tabular}{c|cccc|cccc}
\toprule
\multirow{2}{*}{\textbf{Setting}}
& \multicolumn{4}{c|}{\textbf{\ndonnx}}
& \multicolumn{4}{c}{\textbf{\numba}} \\
\cmidrule(lr){2-5}\cmidrule(lr){6-9}
& \textbf{\pass{1}} & \textbf{\pass{5}} & \textbf{\exec{1}}  & \textbf{\exec{5}}
& \textbf{\pass{1}} & \textbf{\pass{5}} & \textbf{\exec{1}}  & \textbf{\exec{5}} \\
\midrule

\rowcolor{vanillarow}
\baselinename{Vanilla} 
& 23.37 \textcolor[HTML]{909090}{\scriptsize (+0.00)}
& 35.42 \textcolor[HTML]{909090}{\scriptsize (+0.00)}
& 28.52 \textcolor[HTML]{909090}{\scriptsize (+0.00)}
& 43.40 \textcolor[HTML]{909090}{\scriptsize (+0.00)}
& 16.04 \textcolor[HTML]{909090}{\scriptsize (+0.00)}
& 39.51 \textcolor[HTML]{909090}{\scriptsize (+0.00)}
& 26.74 \textcolor[HTML]{909090}{\scriptsize (+0.00)}
& 63.08 \textcolor[HTML]{909090}{\scriptsize (+0.00)} \\

\rowcolor{ragrow}
\baselinename{Direct Synthesis}
& 32.96 \textcolor[HTML]{8B0000}{\scriptsize (+9.59)}
& 40.99 \textcolor[HTML]{8B0000}{\scriptsize (+5.57)}
& 39.11 \textcolor[HTML]{8B0000}{\scriptsize (+10.59)}
& 46.83 \textcolor[HTML]{8B0000}{\scriptsize (+3.43)}
& 17.91 \textcolor[HTML]{8B0000}{\scriptsize (+1.87)}
& 40.69 \textcolor[HTML]{8B0000}{\scriptsize (+1.18)}
& 31.55 \textcolor[HTML]{8B0000}{\scriptsize (+4.81)}
& 66.67 \textcolor[HTML]{8B0000}{\scriptsize (+3.59)} \\

\rowcolor{ragrow}
\baselinename{w/o Graph Pruning}
& 43.85 \textcolor[HTML]{8B0000}{\scriptsize (+20.48)}
& 57.61 \textcolor[HTML]{8B0000}{\scriptsize (+22.19)}
& 52.07 \textcolor[HTML]{8B0000}{\scriptsize (+23.55)}
& 67.36 \textcolor[HTML]{8B0000}{\scriptsize (+23.96)}
& 28.45 \textcolor[HTML]{8B0000}{\scriptsize (+12.41)}
& 46.92 \textcolor[HTML]{8B0000}{\scriptsize (+7.41)}
& 49.25 \textcolor[HTML]{8B0000}{\scriptsize (+22.51)}
& 76.17 \textcolor[HTML]{8B0000}{\scriptsize (+13.09)} \\

\rowcolor{ragrow}
\baselinename{w/o Graph Evolution}
& 44.85 \textcolor[HTML]{8B0000}{\scriptsize (+21.48)}
& 57.66 \textcolor[HTML]{8B0000}{\scriptsize (+22.24)}
& 52.60 \textcolor[HTML]{8B0000}{\scriptsize (+24.08)}
& 67.02 \textcolor[HTML]{8B0000}{\scriptsize (+23.62)}
& 32.57 \textcolor[HTML]{8B0000}{\scriptsize (+16.53)}
& 52.74 \textcolor[HTML]{8B0000}{\scriptsize (+13.23)}
& 53.96 \textcolor[HTML]{8B0000}{\scriptsize (+27.22)}
& 79.16 \textcolor[HTML]{8B0000}{\scriptsize (+16.08)} \\

\rowcolor{newnamerow}
\name
& 47.99 \textcolor[HTML]{8B0000}{\scriptsize (+24.62)}
& 62.76 \textcolor[HTML]{8B0000}{\scriptsize (+27.34)}
& 55.86 \textcolor[HTML]{8B0000}{\scriptsize (+27.34)}
& 71.68 \textcolor[HTML]{8B0000}{\scriptsize (+28.28)}
& 35.61 \textcolor[HTML]{8B0000}{\scriptsize (+19.57)}
& 58.29 \textcolor[HTML]{8B0000}{\scriptsize (+18.78)}
& 56.47 \textcolor[HTML]{8B0000}{\scriptsize (+29.73)}
& 83.60 \textcolor[HTML]{8B0000}{\scriptsize (+20.52)} \\

\bottomrule
\end{tabular}
\end{adjustbox}

\end{table*}
\begin{table}[!tbp]
\centering
\small
\renewcommand{\arraystretch}{1.0}
\setlength{\tabcolsep}{10pt}
\caption{Ablation study of \name on \humaneval under \texttt{numba-cuda} setting. 
}
\label{tab:rq3-humaneval}

\begin{adjustbox}{max width=\linewidth}
\begin{tabular}{ccccc}
\toprule
\bi{Setting} & \textbf{\pass{1}} & \textbf{\pass{5}} & \textbf{\exec{1}} & \textbf{\exec{5}} \\
\midrule

\rowcolor{vanillarow}
\baselinename{Vanilla} 
& 85.43 \textcolor[HTML]{909090}{\scriptsize (-0.00)}
& 92.85 \textcolor[HTML]{909090}{\scriptsize (-0.00)}
& 98.29 \textcolor[HTML]{909090}{\scriptsize (-0.00)}
& 99.95 \textcolor[HTML]{909090}{\scriptsize (-0.00)} \\

\rowcolor{ragrow}
\baselinename{Direct Synthesis}
& 82.50 \textcolor[HTML]{0B3D91}{\scriptsize (-2.93)}
& 88.25 \textcolor[HTML]{0B3D91}{\scriptsize (-4.60)}
& 97.87 \textcolor[HTML]{0B3D91}{\scriptsize (-0.42)}
& 98.78 \textcolor[HTML]{0B3D91}{\scriptsize (-1.17)}  \\

\rowcolor{ragrow}
\baselinename{w/o Graph Pruning}
& 82.13 \textcolor[HTML]{0B3D91}{\scriptsize (-3.30)}
& 89.29 \textcolor[HTML]{0B3D91}{\scriptsize (-3.56)}
& 97.74 \textcolor[HTML]{0B3D91}{\scriptsize (-0.55)}
& 98.64 \textcolor[HTML]{0B3D91}{\scriptsize (-1.31)}  \\

\rowcolor{ragrow}
\baselinename{w/o Graph Evolution}
& 84.63 \textcolor[HTML]{0B3D91}{\scriptsize (-0.80)}
& 90.45 \textcolor[HTML]{0B3D91}{\scriptsize (-2.40)}
& 98.48 \textcolor[HTML]{8B0000}{\scriptsize (+0.19)}
& 99.91 \textcolor[HTML]{0B3D91}{\scriptsize (-0.04)}  \\

\rowcolor{newnamerow}
\name
& 84.02 \textcolor[HTML]{0B3D91}{\scriptsize (-1.41)}
& 90.99 \textcolor[HTML]{0B3D91}{\scriptsize (-1.86)}
& 98.23 \textcolor[HTML]{0B3D91}{\scriptsize (-0.06)}
& 99.78 \textcolor[HTML]{0B3D91}{\scriptsize (-0.17)}  \\
\bottomrule
\end{tabular}
\end{adjustbox}

\end{table}

\vspace{2pt}
\noindent \textbf{Results.}
Table~\ref{tab:ablation} reports the results on \ndonnx and \numba, and Table~\ref{tab:rq3-humaneval} reports the results on \humaneval.

\ding{182} \textit{Both Progressive Graph Evolution and Multidimensional Graph Pruning improve private-library-oriented code generation.}
Removing either component consistently reduces performance on both \ndonnx and \numba.
For example, removing Progressive Graph Evolution decreases \pass{5} on \ndonnx from 62.76\% to 57.66\%.
Removing Multidimensional Graph Pruning causes a larger drop on \numba, where \pass{5} decreases from 58.29\% to 46.92\%.
The degradation is even more pronounced under \baselinename{Direct Synthesis}: on \numba, \exec{1} drops sharply from 56.47\% to 31.55\%.
These results indicate that diversity-oriented evolution and quality-oriented pruning are complementary: the former expands the synthesized data beyond basic API usages, while the latter prevents low-quality samples from being directly learned by the model.

\ding{183} \textit{Multidimensional Graph Pruning helps preserve general code generation capability.}
When Multidimensional Graph Pruning is removed, both \pass{k} and \exec{k} on \humaneval decline.
As shown in Table~\ref{tab:rq3-humaneval}, the full \name achieves 84.02\% \pass{1}, 90.99\% \pass{5}, 98.23\% \exec{1}, and 99.78\% \exec{5}.
After removing Multidimensional Graph Pruning, these scores drop to 82.13\%, 89.29\%, 97.74\%, and 98.64\%, respectively.
These results suggest that unfiltered synthesized data may introduce noisy or incorrect coding patterns, which can negatively affect the model's general code generation capability.
By filtering low-quality samples before fine-tuning, Multidimensional Graph Pruning helps \name improve private-library-oriented code generation while limiting undesired interference with general code generation.

\begin{figure}[!tbp]
    \centering
    \includegraphics[width=0.88\textwidth]{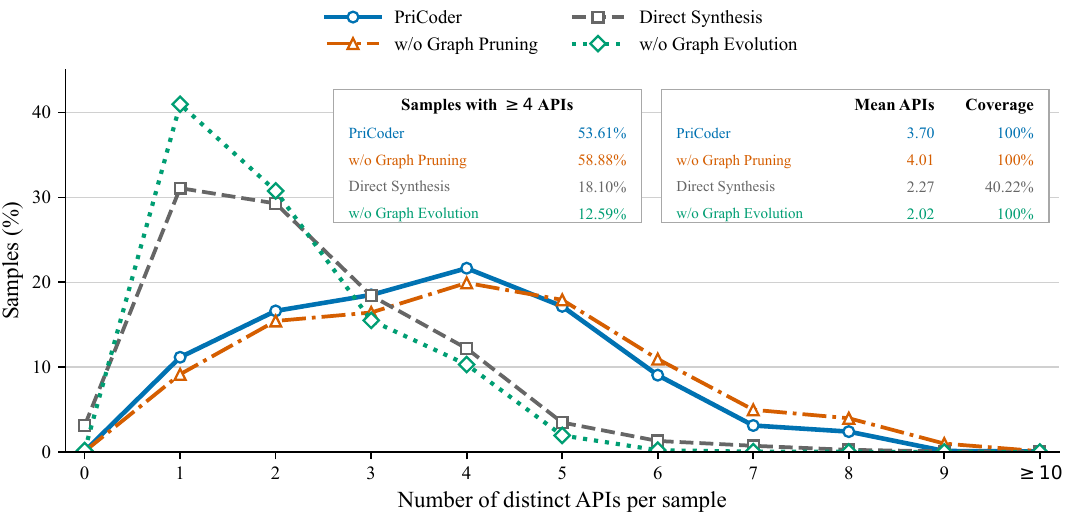}
    \caption{\lrevision{Distribution of the number of distinct \texttt{ndonnx} APIs invoked per synthesized sample. The annotations report the mean number of APIs per sample, the proportion of samples invoking at least four APIs, and API coverage.}}
    \label{fig:rq3-api-distribution}

\end{figure}

\begin{prevision}
\ding{184} \textit{Progressive Graph Evolution increases the diversity of synthesized data.}
As shown in Section~\ref{sec:motivation}, private-library code generation often requires several APIs in one program. Training data with only a few APIs per sample may therefore be insufficient for teaching models to solve such tasks.
We therefore count each distinct \texttt{ndonnx} API once per sample. Figure~\ref{fig:rq3-api-distribution} reports the resulting distributions for the four variants.
Samples generated by \name and \baselinename{w/o Graph Pruning} invoke an average of 3.70 and 4.01 APIs, respectively, and 53.61\% and 58.88\% of them invoke at least four APIs. In contrast, samples generated by \baselinename{Direct Synthesis} and \baselinename{w/o Graph Evolution} invoke only 2.27 and 2.02 APIs on average, and only 18.10\% and 12.59\% of them invoke at least four APIs.
\name, \baselinename{w/o Graph Pruning}, and \baselinename{w/o Graph Evolution} cover all 179 \texttt{ndonnx} APIs, while \baselinename{Direct Synthesis} covers only 72. Although \baselinename{w/o Graph Evolution} achieves full coverage, it averages only 2.02 APIs per sample, showing that API coverage alone does not ensure diversity within individual samples.
These results indicate that Progressive Graph Evolution improves the diversity of synthesized data by increasing the number of APIs used in each sample, thereby providing more varied examples for training models to generate code with private libraries.
\end{prevision}

\begin{boxK}
\small \textbf{Answer to RQ3:}
\name benefits from the complementarity between Progressive Graph Evolution and Multidimensional Graph Pruning: evolution expands the diversity of private-API usage patterns, while pruning filters noisy synthesized samples and helps prevent degradation on general code generation.
\end{boxK}

\subsection{RQ4: How do key factors affect the effectiveness of \name?}
\label{sec:rq4}

\lrevision{Since \name relies on automatically synthesized training data, its effectiveness may depend on the amount of data, how the two evolution stages contribute to the training set, whether the synthesized data transfers across models, and how APIs are sampled during synthesis.
We therefore study four factors: \textit{the scale of synthesized data}, \textit{the model used for data synthesis}, \textit{the allocation between Graph Initial Evolution and Graph Iterative Evolution}, and \textit{the API sampling bias strength $\alpha$}.}

\vspace{2pt}
\noindent \textbf{Setting.}
\lrevision{\ding{182} For the data-scale analysis, we use \deepseek on \ndonnx and vary only the size of the synthesized training set.}
Specifically, we train \deepseek on 500, 1K, 5K, 10K, 20K, 30K, and 40K synthesized samples, while keeping the same synthesis, pruning, and fine-tuning settings as in RQ1.
\lrevision{\ding{183} For the synthesis-model analysis, we use \ndonnx to examine whether the synthesized data must be generated by the same model that will later be fine-tuned.}
We use \qwen and \llama as data synthesizers, each producing 20K training samples, and use the resulting datasets to fine-tune \qwen and \llama.
To further study whether synthesized data transfers across different target-model scales, we control the model family and use data synthesized by \qwen to fine-tune \texttt{Qwen2.5-Coder-3B-Instruct}~\cite{qwencoder}, \qwen, and \texttt{Qwen2.5-Coder-14B-Instruct}~\cite{qwencoder}.
\lrevision{\ding{184} For the evolution-stage analysis, we use \deepseek on both \ndonnx and \numba and vary how the final fine-tuning set is composed of samples generated by Graph Initial Evolution and Graph Iterative Evolution.
We evaluate initial-to-iterative ratios of $\{1{:}0, 3{:}1, 2{:}1, 1{:}1, 1{:}2, 1{:}3, 0{:}1\}$; for example, $1{:}2$ denotes one accepted sample generated through Graph Initial Evolution for every two accepted samples generated through Graph Iterative Evolution in the final fine-tuning set.
Because Graph Iterative Evolution requires existing sample nodes as seeds, the $0{:}1$ setting still generates initial sample nodes to start the iterative process, but excludes them from the final fine-tuning set.
For every ratio, we fix the total number of accepted training samples at 20K for \ndonnx and 6K for \numba, matching RQ1, and keep the pruning and fine-tuning settings unchanged. Thus, only the proportions contributed by the two evolution stages vary.}
\lrevision{\ding{185} To assess sensitivity to the API sampling bias, we use \deepseek on \ndonnx and vary $\alpha$ in Eq.~\ref{eq:api-sampling-weight} over $\{0, 0.1, 0.4, 0.7, 1.0, 10.0\}$, keeping the other settings at their defaults. Here, $\alpha=0$ corresponds to uniform API sampling, while larger values increasingly favor APIs covered by fewer accepted samples.}

\begin{figure}[!tbp]
    \centering
    \includegraphics[width=0.6\textwidth]{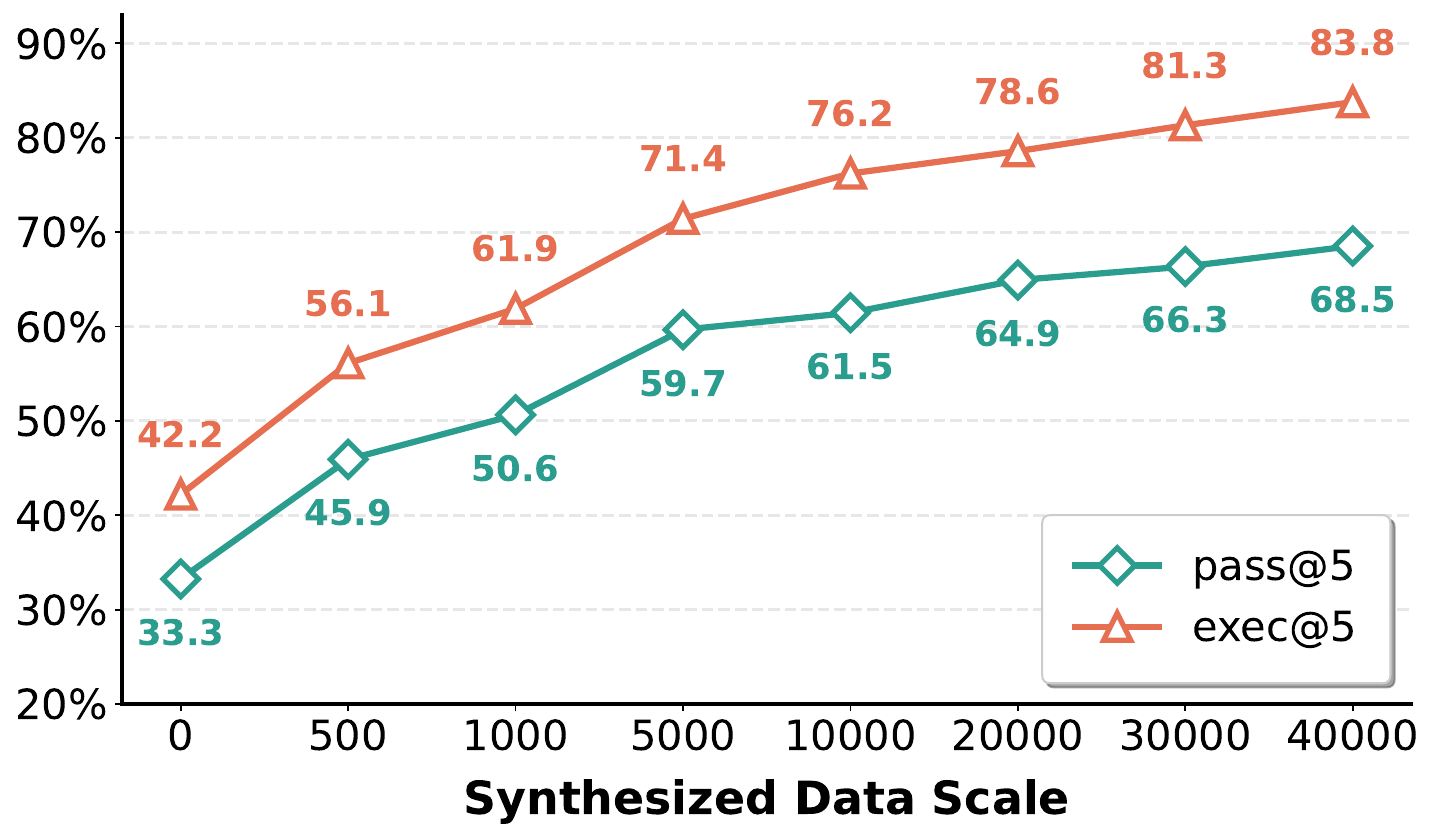}

    \caption{
    \pass{5} and \exec{5} on \ndonnx with different synthesized training data scale.}
    \label{fig:rq4-scale}

\end{figure}

\begin{figure}[!tbp]
    \centering
    \begin{subfigure}[t]{0.4\linewidth}
        \centering
        \includegraphics[width=\linewidth]{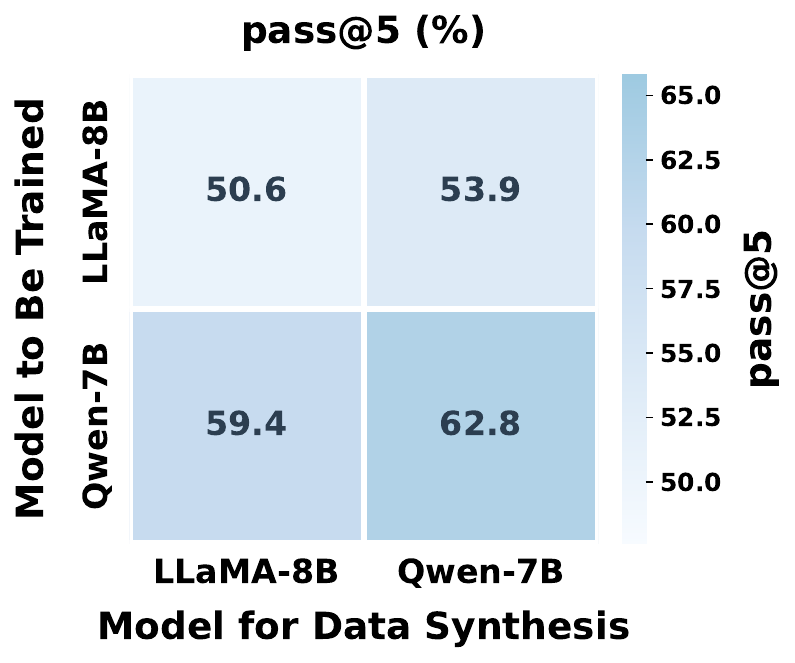}

        \caption{\pass{5} on \ndonnx.}
        \label{fig:rq4-model-pass5}
    \end{subfigure}
    \begin{subfigure}[t]{0.4\linewidth}
        \centering
        \includegraphics[width=\linewidth]{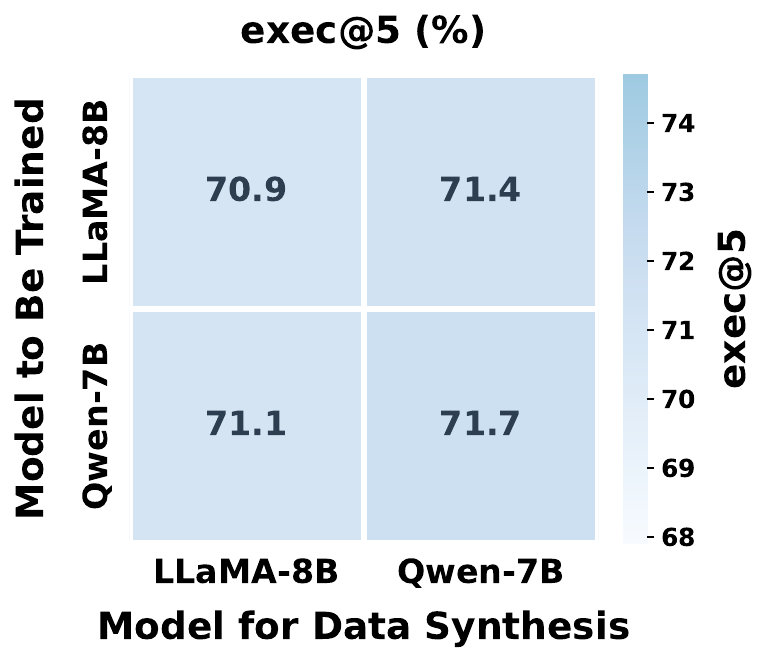}

        \caption{\exec{5} on \ndonnx.}
        \label{fig:rq4-model-exec5}
    \end{subfigure}

    \caption{Impact of the model used for synthesizing training data on the effectiveness of \name. 
    }
    \label{fig:rq4-model}

\end{figure}

\begin{figure*}[!tbp]
    \centering
    \begin{subfigure}[t]{0.48\textwidth}
        \centering
        \includegraphics[width=\linewidth]{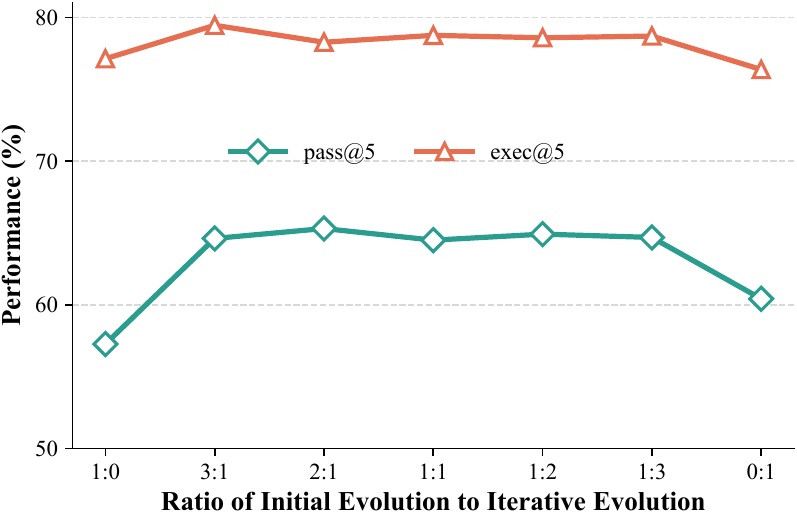}

        \caption{\ndonnx.}
        \label{fig:rq4-evolution-ndonnx}
    \end{subfigure}
    \hfill
    \begin{subfigure}[t]{0.48\textwidth}
        \centering
        \includegraphics[width=\linewidth]{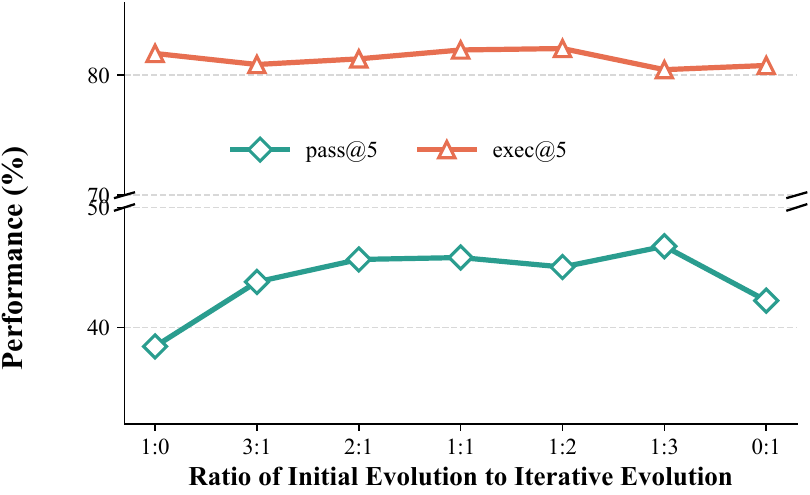}

        \caption{\numba.}
        \label{fig:rq4-evolution-numba}
    \end{subfigure}

    \caption{\lrevision{Impact of the allocation between Graph Initial Evolution and Graph Iterative Evolution.}}
    \label{fig:rq4-evolution-ratio}

\end{figure*}

\begin{figure}[!tbp]
    \centering
    \includegraphics[width=0.6\textwidth]{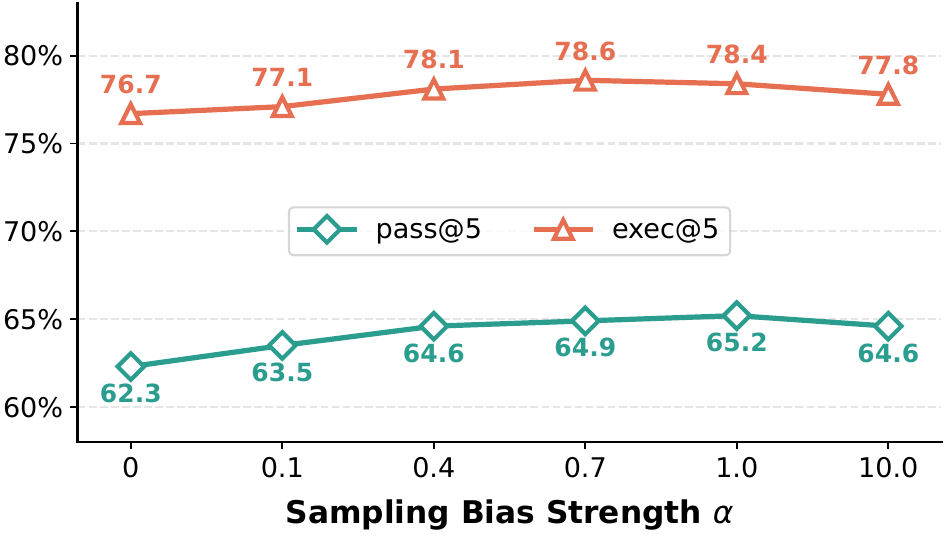}

    \caption{\lrevision{\pass{5} and \exec{5} of \deepseek on \ndonnx with different API sampling bias strengths $\alpha$.}}
    \label{fig:rq4-alpha}

\end{figure}

\vspace{2pt}
\noindent \textbf{Results.}
Figure~\ref{fig:rq4-scale} shows how performance changes with the amount of synthesized data.
Figure~\ref{fig:rq4-model} examines whether synthesized data can transfer across different model families.
Table~\ref{tab:rq4} further studies whether the same synthesized data remains effective across different model scales.
\lrevision{Figure~\ref{fig:rq4-evolution-ratio} reports the effect of varying the allocation between the two evolution stages while holding the total training-data size constant.}
\lrevision{Figure~\ref{fig:rq4-alpha} shows the sensitivity to the API sampling bias strength $\alpha$.}

\bi{Impact of the scale of synthesized data.} 
\ding{182} \textit{Performance improves steadily as synthesized data scales up.}
As shown in Figure~\ref{fig:rq4-scale}, increasing the amount of synthesized data consistently improves both functional correctness and executability.
For example, increasing the data scale from 10K to 40K raises \pass{5} from 61.5\% to 68.5\%.
This suggests that synthesized data scale is an important factor for \name, and organizations with sufficient resources can further improve private-library-oriented code generation by synthesizing more training samples.
\ding{183} \textit{Non-trivial gains can already be achieved with a small training set.}
Even with only 1K synthesized samples, \pass{5} and \exec{5} increase from 33.3\% and 42.2\% to 50.6\% and 61.9\%, respectively.
This indicates that \name does not require an extremely large synthesized dataset to become useful. A modest amount of library-specific data can already provide strong adaptation gains.

\newcommand{\bestcell}[1]{\cellcolor{newnamerow}#1}
\begin{table*}[!tbp]
\centering
\small
\renewcommand{\arraystretch}{1.0}
\setlength{\tabcolsep}{12pt}
\caption{\pass{k} and \exec{k} (\%) on \ndonnx for Qwen models of different scales.
For \baselinename{\name}, all models are trained on synthetic data synthesized by \qwen for \ndonnx.}
\label{tab:rq4}

\begin{adjustbox}{max width=\linewidth}
\begin{tabular}{lccccc}
\toprule
\textbf{Model} & \textbf{Method} & \textbf{\pass{1}} & \textbf{\pass{5}} & \textbf{\exec{1}} & \textbf{\exec{5}} \\
\midrule

\multirow{2}{*}{\texttt{Qwen2.5-Coder-3B-Instruct}}
& \baselinename{Vanilla}
& 15.92 & 29.45 & 21.42 & 37.27 \\
& \cellcolor{newnamerow}\baselinename{\name}
& \cellcolor{newnamerow}41.66
& \cellcolor{newnamerow}58.39
& \cellcolor{newnamerow}52.37
& \cellcolor{newnamerow}73.50 \\
\midrule

\multirow{2}{*}{\texttt{Qwen2.5-Coder-7B-Instruct}}
& \baselinename{Vanilla}
& 23.37 & 35.42 & 28.52 & 43.40 \\
& \cellcolor{newnamerow}\baselinename{\name}
& \cellcolor{newnamerow}47.99
& \cellcolor{newnamerow}62.76
& \cellcolor{newnamerow}55.86
& \cellcolor{newnamerow}71.68 \\
\midrule

\multirow{2}{*}{\texttt{Qwen2.5-Coder-14B-Instruct}}
& \baselinename{Vanilla}
& 24.62 & 36.80 & 29.59 & 44.45 \\
& \cellcolor{newnamerow}\baselinename{\name}
& \cellcolor{newnamerow}53.20
& \cellcolor{newnamerow}66.60
& \cellcolor{newnamerow}60.95
& \cellcolor{newnamerow}73.76 \\
\bottomrule
\end{tabular}
\end{adjustbox}

\end{table*}

\bi{Impact of the model used for data synthesis.}
Figure~\ref{fig:rq4-model} shows that the effectiveness of \name is relatively robust to the model used for data synthesis.
When fine-tuning \llama, using data synthesized by the stronger \qwen achieves performance close to using data synthesized by \llama itself: \pass{5} reaches 53.9\% and 50.6\%, while \exec{5} reaches 71.4\% and 70.9\%, respectively.
Table~\ref{tab:rq4} further shows that the same synthesized dataset can improve Qwen models across different parameter scales.
For example, using data synthesized by \qwen, \name improves \pass{1} from 15.92\% to 41.66\% and \exec{5} from 37.27\% to 73.50\% on \texttt{Qwen2.5-Coder-3B-Instruct}; on the larger \texttt{Qwen2.5-Coder-14B-Instruct}, it improves \pass{1} from 24.62\% to 53.20\% and \pass{5} from 36.80\% to 66.60\%.
These results suggest that synthesized data captures reusable private-library usage patterns rather than model-specific generation habits, and can therefore transfer across both model families and model scales.
In practice, this means that a synthesized dataset may be reused to adapt newly released models, reducing the need to regenerate training data from scratch for every target model.

\begin{prevision}
\bi{Impact of the evolution-stage allocation.}
\ding{182} \textit{Combining the two evolution stages improves functional correctness.}
On both benchmarks, every mixture of initial- and iterative-evolution samples achieves higher \pass{5} than using samples from either stage alone.
Because all variants use the same amount of training data, this improvement cannot be attributed to data scale and instead suggests that samples from the two stages contribute complementary information during fine-tuning.
\ding{183} \textit{The evolution-stage allocation has little effect on executability.}
Unlike \pass{5}, \exec{5} remains stable across the tested ratios on both benchmarks.
We conjecture that this is because executability only requires avoiding runtime failures: verified samples from either stage can teach valid API-call patterns, whereas functional correctness benefits more from the richer API compositions produced by combining both stages.
\ding{184} \textit{The exact mixture ratio is not critical.}
Once both stages are represented, performance remains stable across the mixed ratios, and the default $1{:}2$ ratio stays close to the best-performing configuration across both benchmarks and metrics.
This consistency supports $1{:}2$ as a practical default and shows that the observed benefit does not depend on a narrowly tuned allocation.
\end{prevision}

\begin{prevision}
\bi{Impact of the API sampling bias strength.}
Increasing $\alpha$ from 0 to 0.7 raises \pass{5} from 62.3\% to 64.9\% and \exec{5} from 76.7\% to 78.6\%, showing the benefit of favoring APIs with lower coverage over uniform sampling.
Performance then largely plateaus: across the tested values from $\alpha=0.4$ to $10.0$, \pass{5} and \exec{5} vary by only 0.6 and 0.8 percentage points, respectively.
The default $\alpha=0.7$ achieves the highest \exec{5} and stays within 0.3 percentage points of the best \pass{5}, supporting it as a practical choice without narrow tuning.
\end{prevision}

\begin{boxK}
\small \textbf{Answer to RQ4:}
\lrevision{Increasing synthesized-data scale consistently improves performance, and the resulting data transfers across model families and scales. Under matched data sizes, combining Graph Initial Evolution and Graph Iterative Evolution outperforms using either stage alone in functional correctness and remains robust across different mixed ratios. A moderate API sampling bias improves performance over uniform sampling, with limited sensitivity to further increases in $\alpha$ on \ndonnx.}
\end{boxK}

\section{Discussion}
\label{sec:discussion}

\subsection{Reliability of LLM Judgment}
Multidimensional Graph Pruning relies on LLM-as-a-Judge for overall functionality verification. \lrevision{To assess its reliability, we randomly sampled 100 judge-accepted and 100 judge-rejected samples for manual inspection. Because all inspected samples had already passed syntactic and execution verification, the inspection focused on whether each requirement was realistic and well-defined and whether its solution faithfully implemented the intended functionality. The inspection confirmed that 99 of the 100 judge-accepted samples were valid and that 86 of the 100 judge-rejected samples contained genuine functional or semantic flaws. This asymmetry---only one invalid sample was accepted, whereas 14 valid samples were rejected---suggests that the judge tends to err on the side of rejection.}
Overall, this indicates that the LLM judge provides a useful quality-control signal and helps Multidimensional Graph Pruning remove low-quality samples before fine-tuning.

\begin{table*}[!tbp]
\centering
\small
\renewcommand{\arraystretch}{0.85}
\setlength{\tabcolsep}{4pt}
\caption{Comparison with TreeSynth on \ndonnx and \numba.}
\label{tab:treesynth}

\begin{adjustbox}{max width=\linewidth}
\begin{tabular}{c l | cccc | cccc}
\toprule
\multirow{2}{*}{\textbf{Model}} 
& \multirow{2}{*}{\bi{Method}}
& \multicolumn{4}{c|}{\textbf{\ndonnx}}
& \multicolumn{4}{c}{\textbf{\numba}} \\
\cmidrule(lr){3-6}\cmidrule(lr){7-10}
& 
& \textbf{\pass{1}} & \textbf{\pass{5}} & \textbf{\exec{1}} & \textbf{\exec{5}}
& \textbf{\pass{1}} & \textbf{\pass{5}} & \textbf{\exec{1}} & \textbf{\exec{5}} \\
\midrule

\rowcolor{vanillarow}
\cellcolor{white}
& \baselinename{Vanilla}
& 20.36 & 33.26 & 25.74 & 42.20
& 10.59 & 27.82 & 22.25 & 52.23 \\
\rowcolor{ragrow}
\cellcolor{white}\deepseek
& \baselinename{TreeSynth}
& 35.86 & 50.64 & 40.30 & 59.62
& 14.28 & 33.09 & 30.91 & 65.39 \\
\rowcolor{newnamerow}
\cellcolor{white}
& \name
& 48.17 & 64.92 & 61.72 & 78.58
& 24.12 & 45.04 & 52.73 & 82.21 \\
\midrule

\rowcolor{vanillarow}
\cellcolor{white}
& \baselinename{Vanilla}
& 23.37 & 35.42 & 28.52 & 43.40
& 16.04 & 39.51 & 26.74 & 63.08 \\
\rowcolor{ragrow}
\cellcolor{white}\qwen
& \baselinename{TreeSynth}
& 37.99 & 51.75 & 47.16 & 60.99
& 18.45 & 41.14 & 30.27 & 67.38 \\
\rowcolor{newnamerow}
\cellcolor{white}
& \name
& 47.99 & 62.76 & 55.86 & 71.68
& 35.61 & 58.29 & 56.47 & 83.60 \\
\midrule

\rowcolor{vanillarow}
\cellcolor{white}
& \baselinename{Vanilla}
& 15.86 & 26.55 & 23.14 & 38.65
& 8.13 & 21.73 & 22.78 & 55.25 \\
\rowcolor{ragrow}
\cellcolor{white}\llama
& \baselinename{TreeSynth}
& 19.29 & 32.81 & 33.73 & 43.55
& 9.30 & 25.40 & 24.71 & 60.72 \\
\rowcolor{newnamerow}
\cellcolor{white}
& \name
& 32.31 & 50.57 & 50.18 & 70.92
& 23.58 & 40.84 & 65.94 & 87.90 \\

\bottomrule
\end{tabular}
\end{adjustbox}

\end{table*}

\subsection{Comparison with Data Synthesis Approaches}
Recent studies have explored using LLMs to synthesize training data for improving code generation~\cite{wei2024selfcodealign, wu2025ucoder, li2025api, wang2025epicoder}. However, these approaches are usually designed for standalone programming tasks whose solutions do not depend on an unseen library. In private-library-oriented code generation, synthesized samples must teach the model how to use unfamiliar APIs. Thus, a general synthesis method may generate plausible tasks but provide limited useful supervision.

To assess this issue, we compare \name with \baselinename{TreeSynth}~\cite{wang2025treesynth}, a representative data synthesis method. Its core idea is to ask an LLM to repeatedly turn a broad synthesis request into narrower conditions and then generate samples under these conditions, so that the generated data covers different types of tasks. Since \baselinename{TreeSynth} is designed for requests such as "Synthesize humanEval-style programming problems", which do not specify a private library, it cannot be directly applied to private-library-oriented code generation. We therefore adapt it by using the target API documentation to define what samples should be generated.

We evaluate \baselinename{TreeSynth} and \name on the three models described in Section~\ref{sec:model} and the two private-library benchmarks introduced in Section~\ref{sec:benchmark-construct}. For a fair comparison, \baselinename{TreeSynth} and \name use the same amount of synthesized training data, and all fine-tuning hyperparameters, decoding settings, and evaluation protocols are kept identical.
As shown in Table~\ref{tab:treesynth}, \baselinename{TreeSynth} improves over the vanilla models, but its gains are relatively limited. For example, on \numba with \deepseek, \baselinename{TreeSynth} increases \pass{1} from 10.59\% to 14.28\% and \pass{5} from 27.82\% to 33.09\%, while \name further improves them to 24.12\% and 45.04\%, respectively. Similarly, on \ndonnx with \llama, \baselinename{TreeSynth} improves \pass{1} from 15.86\% to 19.29\%, whereas \name reaches 32.31\%.
This limited improvement suggests that adapting a general synthesis framework with API documentation is still insufficient. \lrevision{In contrast, \name synthesizes samples from private API specifications and filters them through multidimensional verification, making it more suitable for learning private-library API invocation.}

\begin{prevision}
\subsection{Beyond Documentation-Based Retrieval}
\label{sec:richer-retrieval}
Our main experiments follow the predominant documentation-based RAG paradigm for private-library-oriented code generation, which provides retrieved API specifications at inference time~\cite{apifinder, epigen, docprompting, exploracoder}.
However, retrieval can also supply source code or usage examples that expose information absent from API specifications.
To determine whether the limitations observed in our main experiments stem specifically from relying on API specifications, we extend our evaluation to source code and usage examples.
We therefore evaluate two settings that add source code or API-usage examples to the prompt.
\ding{182} \baselinename{Oracle with Source Code} provides both the specifications and source code of the ground-truth required APIs. Because the required APIs and their source files are given directly, this setting excludes retrieval errors.
\ding{183} \baselinename{API Usage Retrieval} uses BGE-M3~\cite{multi2024m3} to retrieve the four most similar requirement--solution pairs from the same model-specific $\mathcal{D}_{\mathrm{syn}}$ used to fine-tune \name and adds these four examples to the prompt.
We evaluate both settings with \qwen and \llama on \ndonnx and \numba under the evaluation settings of RQ1.
The results are reported in Table~\ref{tab:dis-richer-context}.

\begin{table*}[!tbp]
\crevision
\centering
\small
\renewcommand{\arraystretch}{0.95}
\setlength{\tabcolsep}{3.5pt}
\caption{\lrevision{Comparison of \name with methods that provide source code or retrieved API-usage examples at inference time on \ndonnx and \numba.}}
\label{tab:dis-richer-context}

\begin{adjustbox}{max width=\linewidth}
\begin{tabular}{c >{\raggedright\arraybackslash}p{3.2cm} | cccc | cccc}
\toprule
\multirow{2}{*}{\textbf{Model}}
& \multirow{2}{*}{\textbf{Method}}
& \multicolumn{4}{c|}{\textbf{\ndonnx}}
& \multicolumn{4}{c}{\textbf{\numba}} \\
\cmidrule(lr){3-6}\cmidrule(lr){7-10}
&
& \textbf{\pass{1}} & \textbf{\pass{5}} & \textbf{\exec{1}} & \textbf{\exec{5}}
& \textbf{\pass{1}} & \textbf{\pass{5}} & \textbf{\exec{1}} & \textbf{\exec{5}} \\
\midrule

\multirow{5}{*}{\qwen}
& \baselinename{Vanilla}
& 23.37 & 35.42 & 28.52 & 43.40
& 16.04 & 39.51 & 26.74 & 63.08 \\
& \baselinename{Oracle}
& 52.60 & \textbf{66.16} & 60.24 & \textbf{73.87}
& 24.97 & 50.54 & 35.78 & 71.36 \\
& \baselinename{Oracle with Source Code}
& \textbf{53.31} & 64.82 & \textbf{61.18} & 73.51
& 5.08 & 17.23 & 10.37 & 32.93 \\
& \baselinename{API Usage Retrieval}
& 41.89 & 52.76 & 49.82 & 62.52
& 30.96 & 46.77 & 53.10 & 74.82 \\
& \name
& 47.99 & 62.76 & 55.86 & 71.68
& \textbf{35.61} & \textbf{58.29} & \textbf{56.47} & \textbf{83.60} \\
\midrule

\multirow{5}{*}{\llama}
& \baselinename{Vanilla}
& 15.86 & 26.55 & 23.14 & 38.65
& 8.13 & 21.73 & 22.78 & 55.25 \\
& \baselinename{Oracle}
& 30.36 & 47.45 & 46.57 & 70.38
& 13.10 & 33.66 & 29.36 & 62.55 \\
& \baselinename{Oracle with Source Code}
& 29.76 & 47.52 & 48.22 & 69.87
& 5.13 & 14.05 & 13.69 & 39.21 \\
& \baselinename{API Usage Retrieval}
& 22.13 & 35.06 & 33.14 & 49.09
& 17.17 & 28.31 & 50.43 & 71.75 \\
& \name
& \textbf{32.31} & \textbf{50.57} & \textbf{50.18} & \textbf{70.92}
& \textbf{23.58} & \textbf{40.84} & \textbf{65.94} & \textbf{87.90} \\

\bottomrule
\end{tabular}
\end{adjustbox}

\end{table*}

Providing the source code of the ground-truth required APIs neither closes the performance gap with \name nor consistently improves \baselinename{Oracle}.
\ding{182} \name remains more effective than \baselinename{Oracle with Source Code}. Specifically, \baselinename{Oracle with Source Code} falls below \name on 12 of the 16 reported metrics, even though \name receives no API specifications, source code, or usage examples at inference time. Moreover, \baselinename{Oracle with Source Code} is evaluated under idealized conditions: it receives the specifications and source code of exactly the required APIs without retrieval errors. In practice, systems must first retrieve the relevant APIs and source files and therefore face potential retrieval errors.
\ding{183} Adding the source code of the ground-truth required APIs does not consistently improve \baselinename{Oracle}. For example, it improves only 4 of the 16 metrics and degrades the remaining 12.
Retrieving API-usage examples consistently improves the vanilla models but remains less effective than \name.
\ding{182} Retrieved API-usage examples provide useful guidance without parameter updates. \baselinename{API Usage Retrieval} improves over \baselinename{Vanilla} on all 16 metrics. For example, on \ndonnx with \qwen, it increases \pass{1} from 23.37\% to 41.89\% and \exec{1} from 28.52\% to 49.82\%.
\ding{183} \name remains more effective than retrieving API-usage examples at inference time. \name outperforms \baselinename{API Usage Retrieval} on all 16 metrics. For example, on \numba with \llama, \name achieves 23.58\% \pass{1} and 65.94\% \exec{1}, compared with 17.17\% and 50.43\%, respectively.

Together, these results show that the limitations of retrieval-based prompting are not confined to mainstream API specifications: neither providing the source code of the required APIs nor retrieving API-usage examples consistently matches \name.
This evidence further supports \name's approach of learning how APIs are used during fine-tuning rather than relying solely on additional API information supplied in the prompt.
\end{prevision}

\begin{prevision}
\subsection{Fine-Tuning on Library Source Code}
Private-library owners typically have access to the library source code.
Because this code is readily available, we examine whether it can serve as an alternative source of supervision for learning API invocation.
We construct \baselinename{Source-SFT} by extracting API signature--implementation pairs from each library and fine-tuning \qwen and \llama with the same training configuration as \name.
\seclrevision{The training examples follow an instruction--response format, where the instruction specifies the library, fully qualified API name, signature, and docstring, and the target response is the corresponding Python implementation extracted from the library source code.}
We evaluate the models on \ndonnx and \numba using the same inference settings as RQ1.

\begin{table}[!tbp]
\crevision
\centering
\small
\renewcommand{\arraystretch}{0.95}
\setlength{\tabcolsep}{5pt}
\caption{\lrevision{Comparison of fine-tuning on library source code with \baselinename{Vanilla} and \name on \ndonnx and \numba. All values are percentages.}}
\label{tab:dis-source-sft}

\begin{adjustbox}{max width=\linewidth}
\begin{tabular}{c l | cccc | cccc}
\toprule
\multirow{2}{*}{\textbf{Model}}
& \multirow{2}{*}{\textbf{Method}}
& \multicolumn{4}{c|}{\textbf{\ndonnx}}
& \multicolumn{4}{c}{\textbf{\numba}} \\
\cmidrule(lr){3-6}\cmidrule(lr){7-10}
&
& \textbf{\pass{1}} & \textbf{\pass{5}} & \textbf{\exec{1}} & \textbf{\exec{5}}
& \textbf{\pass{1}} & \textbf{\pass{5}} & \textbf{\exec{1}} & \textbf{\exec{5}} \\
\midrule

\multirow{3}{*}{\qwen}
& \baselinename{Vanilla}
& 23.37 & 35.42 & 28.52 & 43.40
& 16.04 & 39.51 & 26.74 & 63.08 \\
& \baselinename{Source-SFT}
& 19.76 & 27.71 & 21.95 & 30.34
& 2.41 & 9.11 & 5.19 & 18.16 \\
& \name
& \textbf{47.99} & \textbf{62.76} & \textbf{55.86} & \textbf{71.68}
& \textbf{35.61} & \textbf{58.29} & \textbf{56.47} & \textbf{83.60} \\
\midrule

\multirow{3}{*}{\llama}
& \baselinename{Vanilla}
& 15.86 & 26.55 & 23.14 & 38.65
& 8.13 & 21.73 & 22.78 & 55.25 \\
& \baselinename{Source-SFT}
& 5.56 & 7.76 & 6.27 & 10.27
& 1.34 & 3.60 & 2.83 & 9.46 \\
& \name
& \textbf{32.31} & \textbf{50.57} & \textbf{50.18} & \textbf{70.92}
& \textbf{23.58} & \textbf{40.84} & \textbf{65.94} & \textbf{87.90} \\

\bottomrule
\end{tabular}
\end{adjustbox}

\end{table}

Table~\ref{tab:dis-source-sft} shows that \baselinename{Source-SFT} underperforms even \baselinename{Vanilla} in every reported setting.
For example, on \numba with \qwen, \pass{1} decreases from 16.04\% to 2.41\% and \exec{1} from 26.74\% to 5.19\%, whereas \name reaches 35.61\% and 56.47\%, respectively.
These results indicate that private-library source code alone is not effective training data for improving a model's ability to use the library.
\end{prevision}

\begin{prevision}
\subsection{Distinguishing API Learning from Format Adaptation}
\label{sec:cross-library-transfer}
One potential concern is that \name's gains may arise from adapting the model to the task format rather than from learning API usage. \ndonnx and \numba allow us to separate these two effects: their tasks follow comparable formats, but their solutions require APIs from different libraries. We therefore conduct a cross-library transfer experiment with \qwen and \llama, fine-tuning each model on the synthesized data constructed for one library and evaluating it on the other under the same settings as RQ1. In this setting, the models are fine-tuned on tasks with a comparable format, but not on examples that use the APIs required by the evaluation benchmark. If format adaptation were responsible for the gains, the transferred models should consistently outperform \baselinename{Vanilla}. The results in Table~\ref{tab:dis-cross-library-transfer} show that they do not: cross-library transfer performs similarly to or worse than \baselinename{Vanilla} across both models and benchmarks, while remaining substantially below \name trained on data for the corresponding library. For example, on \ndonnx with \qwen, cross-library transfer achieves 23.55\% \pass{1}, nearly identical to the 23.37\% of \baselinename{Vanilla}, whereas \name achieves 47.99\%. Thus, these results confirm that \name improves the model's ability to use the target library, rather than merely adapting it to the task format.

\begin{table*}[!tbp]
\crevision
\centering
\small
\renewcommand{\arraystretch}{0.95}
\caption{\lrevision{Cross-library transfer results in \pass{k} (\%) on \ndonnx and \numba. For \baselinename{Cross-Library Transfer}, each model is fine-tuned on the synthesized data constructed for the other library.}}
\label{tab:dis-cross-library-transfer}

\begin{tabular*}{\textwidth}{@{\extracolsep{\fill}}c l | ccc | ccc@{}}
\toprule
\multirow{2}{*}{\textbf{Model}}
& \multirow{2}{*}{\textbf{Method}}
& \multicolumn{3}{c|}{\textbf{\ndonnx}}
& \multicolumn{3}{c}{\textbf{\numba}} \\
\cmidrule(lr){3-5}\cmidrule(lr){6-8}
&
& \textbf{\pass{1}} & \textbf{\pass{3}} & \textbf{\pass{5}}
& \textbf{\pass{1}} & \textbf{\pass{3}} & \textbf{\pass{5}} \\
\midrule

\multirow{3}{*}{\qwen}
& \baselinename{Vanilla}
& 23.37 & 31.69 & 35.42
& 16.04 & 31.82 & 39.51 \\
& \baselinename{Cross-Library Transfer}
& 23.55 & 31.90 & 34.71
& 12.57 & 25.38 & 33.21 \\
& \name
& 47.99 & 59.25 & 62.76
& 35.61 & 51.83 & 58.29 \\
\midrule

\multirow{3}{*}{\llama}
& \baselinename{Vanilla}
& 15.86 & 23.50 & 26.55
& 8.13 & 16.97 & 21.73 \\
& \baselinename{Cross-Library Transfer}
& 12.66 & 20.79 & 24.29
& 6.26 & 14.51 & 20.19 \\
& \name
& 32.31 & 45.28 & 50.57
& 23.58 & 35.34 & 40.84 \\

\bottomrule
\end{tabular*}

\end{table*}

\end{prevision}

\begin{prevision}
\subsection{Impact of Train--Test Similarity}
\label{sec:train-test-similarity}
The performance gains of \name may partly stem from synthesized training samples that resemble benchmark instances. We therefore examine train--test similarity and its impact on model performance.

\bi{Train--test similarity.}
To determine how closely synthesized training samples resemble benchmark instances, we analyze the datasets synthesized by all three models for \ndonnx (20K samples each) and \numba (6K each). Using \texttt{Qwen3-Embedding-4B}~\cite{zhang2025qwen3embedding}, we compute cosine similarity between training and benchmark requirements and between their solutions, retaining the maximum for each training sample at each level. For API composition, we check whether each training solution invokes exactly the same set of APIs as any benchmark solution.
Table~\ref{tab:dis-train-test-similarity} shows that only 0.03--1.15\% of training samples share an identical API set with any benchmark instance, indicating limited exact overlap in API compositions. Figure~\ref{fig:dis-train-test-similarity} further shows that maximum requirement and solution similarities are concentrated away from 1.0 across all three models and both libraries. Together, these results suggest that the synthesized training samples are generally distinct from benchmark instances in their requirements, solutions, and API compositions.

\begin{table}[!tbp]
\crevision
\centering
\small
\caption{\lrevision{Mean train--test similarity for each synthesized dataset. Requirement and solution columns report mean maximum cosine similarity; API exact match reports the percentage of training samples whose API set matches at least one benchmark instance.}}
\label{tab:dis-train-test-similarity}
\begin{adjustbox}{max width=\linewidth}
\begin{tabular}{llccc}
\toprule
\textbf{Benchmark} & \textbf{Synthesis model} & \textbf{Requirement} & \textbf{Solution} & \textbf{API exact match (\%)} \\
\midrule
\multirow{3}{*}{\ndonnx}
& \qwen & 0.608 & 0.680 & 0.57 \\
& \llama & 0.609 & 0.666 & 0.36 \\
& \deepseek & 0.608 & 0.675 & 1.15 \\
\midrule
\multirow{3}{*}{\numba}
& \qwen & 0.718 & 0.785 & 0.25 \\
& \llama & 0.718 & 0.736 & 0.25 \\
& \deepseek & 0.702 & 0.744 & 0.03 \\
\bottomrule
\end{tabular}
\end{adjustbox}
\end{table}

\begin{figure*}[!tbp]
    \centering
    \includegraphics[width=0.82\textwidth]{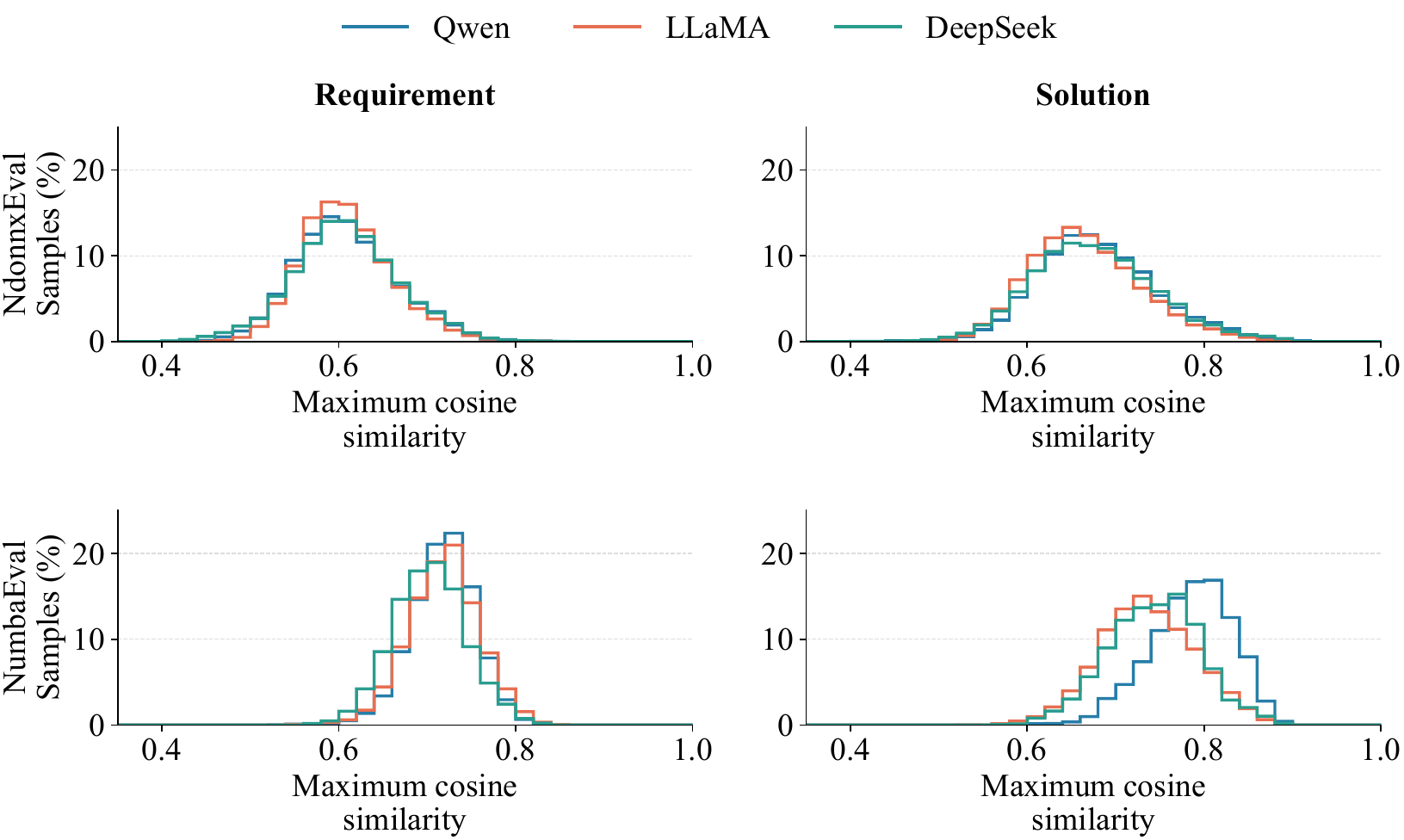}
    \caption{\lrevision{Train--test similarity distributions on \ndonnx (top) and \numba (bottom). Columns show requirement (left) and solution (right).}}
    \label{fig:dis-train-test-similarity}
    \Description{Four panels compare three synthesis models on two libraries. Rows show Ndonnx and Numba; columns show maximum cosine similarity histograms for requirements and solutions.}
\end{figure*}

\bi{Removing similar training samples.}
We next examine whether the performance gains depend on the training samples most similar to benchmark instances. For \qwen and \llama on \ndonnx, we remove samples with an exact API-set match or ranked in the top 5\% by requirement or solution similarity. As a baseline, we randomly remove the same number of samples from the original training dataset. This comparison distinguishes the effect of removing similar samples from that of simply reducing training-data size. We fine-tune each model on the two resulting datasets and evaluate their performance.
Table~\ref{tab:dis-similarity-filtering} shows that removing the most similar samples yields nearly the same effectiveness as random deletion, with differences of at most 0.18 and 0.62 percentage points in \pass{1} and \pass{5}, respectively. These results indicate that \name's performance gains do not rely heavily on training samples that closely resemble benchmark instances.

\begin{table*}[!tbp]
\crevision
\centering
\small
\setlength{\tabcolsep}{7pt}
\caption{\lrevision{Performance (\%) on \ndonnx after similarity filtering and random deletion.}}
\label{tab:dis-similarity-filtering}
\begin{adjustbox}{max width=\linewidth}
\begin{tabular}{llrrrrr}
\toprule
\textbf{Model} & \textbf{Condition} & \textbf{Retained}
& \textbf{\pass{1}} & \textbf{\pass{5}} & \textbf{\exec{1}} & \textbf{\exec{5}} \\
\midrule
\multirow{2}{*}{\qwen}
& Similarity filtering & 18,346 & 46.63 & 57.82 & 54.32 & 65.84 \\
& Random deletion & 18,346 & 46.75 & 58.44 & 54.67 & 65.10 \\
\midrule
\multirow{2}{*}{\llama}
& Similarity filtering & 18,432 & 32.25 & 49.52 & 48.93 & 70.50 \\
& Random deletion & 18,432 & 32.43 & 49.58 & 47.87 & 67.97 \\
\bottomrule
\end{tabular}
\end{adjustbox}
\end{table*}

\bi{Data scale after similarity filtering.}
To assess whether increasing training-data size remains beneficial after removing the most similar samples, we repeat the RQ4 data-scale experiment with \deepseek on \ndonnx, applying both deletion strategies at each scale while keeping the other settings unchanged. Both \pass{5} and \exec{5} increase consistently with data size (Figure~\ref{fig:dis-similarity-scale}). Similarity filtering and random deletion yield nearly identical performance across all data scales. These results suggest that \name learns reusable API-usage knowledge rather than relying solely on highly similar training samples.

\begin{figure*}[!tbp]
    \centering
    \includegraphics[width=\textwidth]{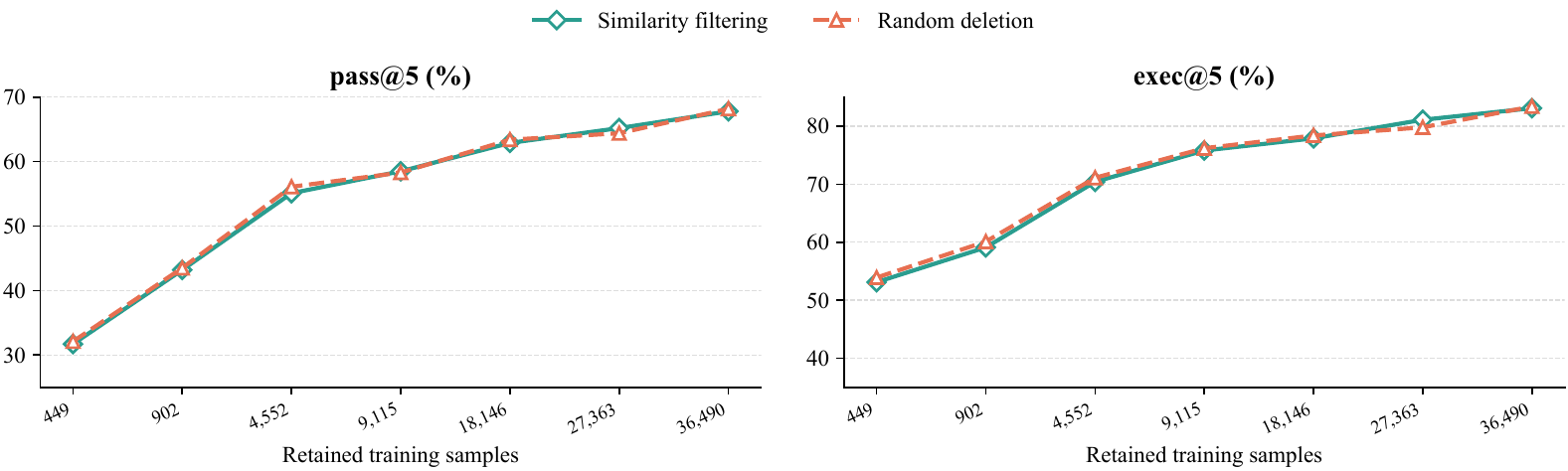}
    \caption{\lrevision{Effect of synthesized training-data scale after similarity filtering on \ndonnx with \deepseek. Random deletion retains the same number of samples at each size.}}
    \label{fig:dis-similarity-scale}
    \Description{Two line charts compare pass at 5 and exec at 5 after similarity filtering and size-matched random deletion. Both metrics increase consistently with the number of retained training samples under both conditions.}
\end{figure*}

\end{prevision}

\begin{prevision}
\subsection{Assessing Model Familiarity with the Target Libraries}
\label{sec:prior-exposure}
Evaluating private-library-oriented code generation requires target libraries that the models have not already learned. To verify this condition directly rather than infer it from release dates, we apply a zero-shot exact-signature-recall probe to the three models evaluated in Section~\ref{sec:exp}. For every \texttt{ndonnx} and \texttt{numba-cuda} API with a non-empty reference signature, we provide only the library and API names and ask the model to generate the signature. We query each API ten times at a temperature of 0.5 and top-$p$ of 0.95, counting only exact matches to the reference signature as correct. Here, \recall{k} measures the probability that at least one of $k$ generated signatures exactly matches the reference signature, averaged over the evaluated APIs in each library. We report \recall{1} and \recall{5} using the same unbiased estimator as \pass{k} (Section~\ref{sec:metric}), treating each API as an evaluation instance and each exact signature match as a success.

\begin{table}[!tbp]
\crevision
\centering
\small
\renewcommand{\arraystretch}{1.02}
\setlength{\tabcolsep}{10pt}
\caption{\lrevision{Zero-shot exact-signature recall (\%) on the target libraries.}}
\label{tab:signature-recall}

\begin{adjustbox}{max width=\linewidth}
\begin{tabular}{llcc}
\toprule
\textbf{Model} & \textbf{Library} & \textbf{\recall{1}} & \textbf{\recall{5}} \\
\midrule
\multirow{2}{*}{\deepseek}
& \texttt{ndonnx} & 0.22 & 1.12 \\
& \texttt{numba-cuda} & 2.99 & 6.12 \\
\midrule
\multirow{2}{*}{\qwen}
& \texttt{ndonnx} & 0.22 & 1.12 \\
& \texttt{numba-cuda} & 7.32 & 11.55 \\
\midrule
\multirow{2}{*}{\llama}
& \texttt{ndonnx} & 0.00 & 0.00 \\
& \texttt{numba-cuda} & 0.17 & 0.57 \\
\bottomrule
\end{tabular}
\end{adjustbox}

\end{table}

Table~\ref{tab:signature-recall} reports uniformly low recall. On \texttt{ndonnx}, the highest \recall{1} and \recall{5} are only 0.22\% and 1.12\%, respectively. Recall is higher on \texttt{numba-cuda}, but the highest \recall{1} and \recall{5} remain only 7.32\% and 11.55\%. Our case analysis suggests that this difference may arise because \texttt{numba-cuda} contains many simple APIs, including APIs with only a single basic parameter, whose signatures may be guessed without prior knowledge of the library. Overall, the models cannot reliably recall the target API signatures without documentation and have little usable prior knowledge of either library. This evidence supports using both libraries as proxies for private libraries in our experiments.
\end{prevision}

\begin{prevision}
\subsection{Variance Across Training Runs}
\label{sec:training-variance}
Potential variation in data synthesis and fine-tuning may affect the reliability of our experimental results.
To assess this variation, we repeat the complete data synthesis and fine-tuning process of \name four times for each of \qwen, \llama, and \deepseek on \ndonnx.
Each run uses a different random seed to generate a new training dataset and fine-tune the model, with the same data size and experimental settings as RQ1.
Figure~\ref{fig:dis-training-variance} summarizes the means and standard deviations across these four repeated runs.
The standard deviations range from 0.33 to 1.49 percentage points across all models and metrics, indicating that \name exhibits limited variation across runs.
The mean scores remain substantially above \baselinename{Vanilla}. For example, \qwen achieves a mean \pass{1} of 47.8\%, compared with 23.37\% for \baselinename{Vanilla}.
These results support the reliability of \name's improvements on \ndonnx across repeated data synthesis and fine-tuning.

\begin{figure*}[!tbp]
    \centering
    \includegraphics[width=0.84\textwidth]{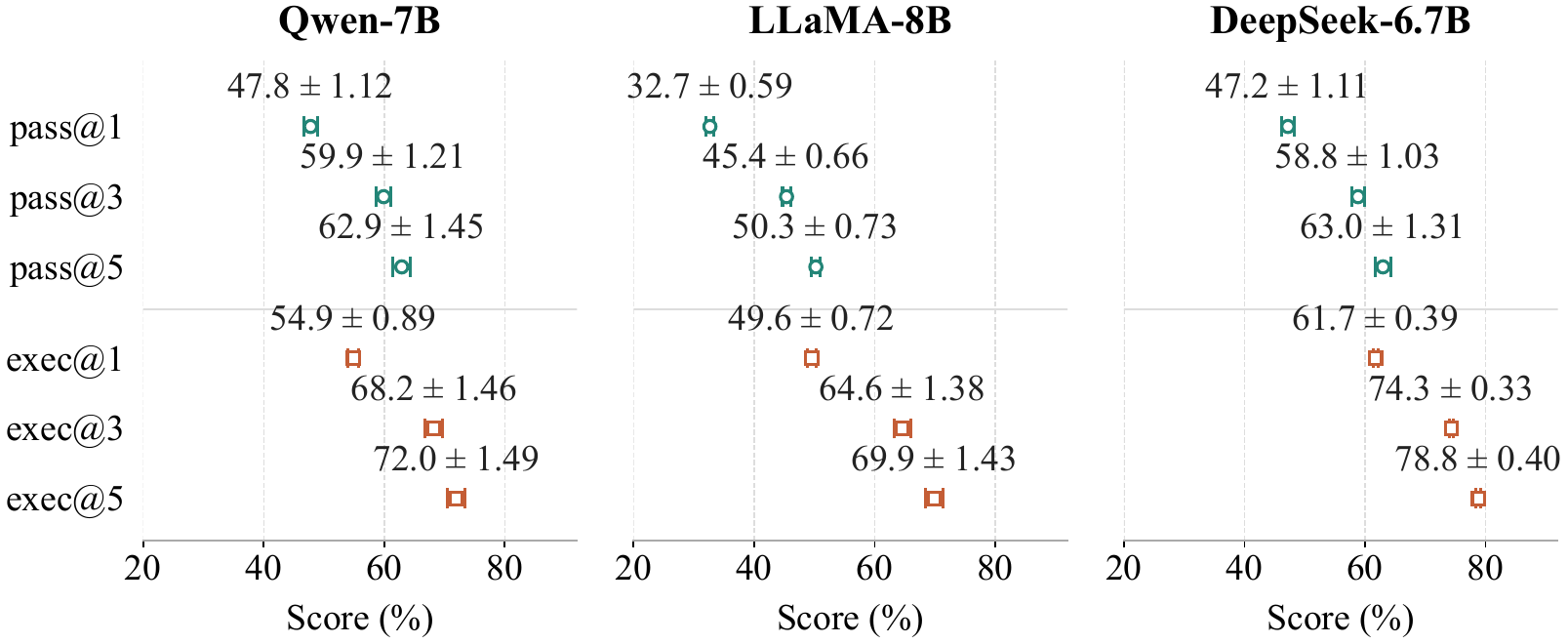}
    \caption{\lrevision{Performance (\%) of \name on \ndonnx across four runs of data synthesis and fine-tuning per model. Points show means, and error bars indicate standard deviations.}}
    \label{fig:dis-training-variance}
\end{figure*}

\end{prevision}

\begin{prevision}
\subsection{Cost of \name}
\label{sec:cost}
To assess the cost of \name, we measure the wall-clock time required to synthesize 20K training samples for \texttt{ndonnx} and fine-tune each of the three models in Section~\ref{sec:model}. We use a server equipped with eight NVIDIA A100-SXM4-80GB GPUs, allocating eight GPUs to data synthesis and two to fine-tuning, with the remaining settings matching RQ1. Table~\ref{tab:dis-cost} reports the time for each stage and the total GPU-hours. Data synthesis takes 15.4--27.2 hours, whereas fine-tuning takes only 33--42 minutes. The complete process therefore takes roughly one day (16.0--27.8 hours), representing a manageable offline adaptation cost that can be amortized over subsequent code-generation requests. We also observe that stronger base models complete data synthesis faster. For example, \qwen, which has the highest Vanilla \pass{1} on \ndonnx among the three models, requires the least synthesis time. This trend suggests that advances in base-model capability may further improve the synthesis efficiency of \name.

\begin{table}[!tbp]
\crevision
\centering
\small
\renewcommand{\arraystretch}{1.0}
\setlength{\tabcolsep}{9pt}
\caption{\lrevision{Cost of \name for synthesizing 20K \texttt{ndonnx} training samples and fine-tuning each model. Times are reported as hh:mm:ss. GPU-hours are calculated as eight times the synthesis duration plus twice the fine-tuning duration, in hours.}}
\label{tab:dis-cost}
\begin{adjustbox}{max width=\linewidth}
\begin{tabular}{lcccc}
\toprule
\textbf{Model} & \makecell{\textbf{Synthesis}\\(8 GPUs)} & \makecell{\textbf{Fine-tuning}\\(2 GPUs)} & \textbf{Total time} & \textbf{GPU-hours} \\
\midrule
\qwen     & 15:26:23 & 00:33:17 & 15:59:40 & 124.63 \\
\deepseek & 22:57:22 & 00:34:44 & 23:32:06 & 184.81 \\
\llama    & 27:09:15 & 00:41:26 & 27:50:41 & 218.61 \\
\bottomrule
\end{tabular}
\end{adjustbox}

\end{table}

\end{prevision}

\begin{secprevision}
\subsection{Analysis of Multidimensional Graph Pruning}
\label{sec:verification-analysis}
Multidimensional Graph Pruning filters synthesized samples before they are used for fine-tuning.
To understand how this filtering works in practice, we examine where samples are rejected and how much of the synthesized code is exercised by the generated tests.
We conduct this analysis on \texttt{ndonnx} using \qwen, \deepseek, and \llama.
Table~\ref{tab:verification-analysis} reports the acceptance rate of each validator and the average line and branch coverage of the tests generated during execution verification.
Each acceptance rate is calculated relative to the samples entering that stage.

\begin{table}[!tbp]
\seccrevision
\centering
\small
\renewcommand{\arraystretch}{1.0}
\setlength{\tabcolsep}{7pt}
\caption{\seclrevision{Acceptance rates of the three verification stages and average coverage of the generated tests on \texttt{ndonnx} (\%).}}
\label{tab:verification-analysis}

\begin{adjustbox}{max width=\linewidth}
\begin{tabular}{lccccc}
\toprule
\multirow{2}{*}{\textbf{Model}} & \multicolumn{3}{c}{\textbf{Acceptance rate}} & \multicolumn{2}{c}{\textbf{Test coverage}} \\
\cmidrule(lr){2-4}\cmidrule(lr){5-6}
& \textbf{Syntactic} & \textbf{Execution} & \textbf{Functionality} & \textbf{Line} & \textbf{Branch} \\
\midrule
\qwen     & 97.89 & 43.74 & 97.62 & 55.94 & 80.69 \\
\deepseek & 85.58 & 14.91 & 98.02 & 59.44 & 75.23 \\
\llama    & 82.16 & 21.30 & 96.01 & 57.26 & 76.01 \\
\bottomrule
\end{tabular}
\end{adjustbox}
\end{table}

Most samples pass syntactic verification, but only 14.91--43.74\% of these samples pass execution verification.
Among the samples that survive both checks, 96.01--98.02\% also pass functionality verification.
The generated tests achieve average line coverage of 55.94--59.44\% and branch coverage of 75.23--80.69\%.
The acceptance rates show that execution verification performs most of the filtering, removing samples that pass syntax checks but fail the generated tests.
\end{secprevision}

\subsection{Threats to Validity}
\label{sec:threats}

\noindent
\ding{182} \textit{Validity of private-library proxies.}
Ideally, private-library-oriented code generation should be evaluated on real enterprise libraries.
However, such libraries are usually closed-source and often tied to proprietary execution environments.
Therefore, following the practice of prior work~\cite{apifinder, exploracoder}, we construct benchmarks from recently released public libraries that are unfamiliar to the evaluated models.
\lrevision{As shown by the zero-shot exact-signature-recall results in Table~\ref{tab:signature-recall}, the evaluated models have little usable prior knowledge of their APIs, supporting the use of these libraries as proxies for private libraries.}

\noindent
\ding{183} \textit{Quality of synthesized training data.}
Another threat lies in the quality of the synthesized requirement-solution pairs used for training.
Although \name applies Multidimensional Graph Pruning to filter samples through syntactic verification, execution verification, and functionality verification, these checks cannot fully guarantee semantic correctness.
In particular, LLM-generated tests may miss corner cases, and LLM-based judgment may still fail to identify subtle misuse of unfamiliar APIs.
To reduce this threat, we combine multiple complementary validators and manually inspect both accepted and rejected samples after pruning.
The manual inspection shows that the accepted samples are highly reliable and most rejected samples are indeed flawed, which provides empirical evidence for the effectiveness of Multidimensional Graph Pruning.

\noindent
\ding{184} \textit{Generalizability of our results.}
A potential threat is whether our findings generalize across tasks and models.
To mitigate this threat, we evaluate \name on multiple benchmarks and models.
Following prior work~\cite{chen2022codet, athiwaratkun2022multi, li2025structured}, we report unbiased \pass{k} and \exec{k}, sample 10 candidate solutions per instance to reduce randomness, and compare \name with three representative prior approaches as well as an \baselinename{Oracle} baseline that directly provides the required API specifications.

\begin{prevision}
\noindent
\ding{185} \textit{Scope of the RAG evaluation.}
Our main evaluation focuses on documentation-based RAG, where retrieved API specifications are provided to the model at inference time.
This setting follows the predominant paradigm in prior work on private-library-oriented code generation~\cite{apifinder, epigen, docprompting, exploracoder} and reflects the practical scarcity of API-usage examples for private libraries~\cite{epigen, majumdar2025genetic}.
To examine whether source code and API-usage examples can improve generation, Section~\ref{sec:richer-retrieval} evaluates both providing the source code of the ground-truth required APIs and retrieving relevant requirement--solution pairs as API-usage examples.
As shown in Table~\ref{tab:dis-richer-context}, \baselinename{API Usage Retrieval} consistently improves over \baselinename{Vanilla} but remains less effective than \name across the evaluated models and benchmarks.
Comparing the same table with Table~\ref{tab:rq1} further shows that \baselinename{Oracle with Source Code} is lower than \name on 12 of the 16 reported metrics, despite receiving both the specifications and source code of the ground-truth required APIs.
These additional results extend our analysis beyond documentation-only retrieval and further support our main finding that, in the evaluated settings, providing API information at inference time does not consistently resolve the difficulty of private-library API invocation.
\end{prevision}

\section{Conclusion}
\label{sec:conclusion}

In this paper, we studied private-library-oriented code generation, a practical yet challenging setting where LLMs must generate code that correctly invokes APIs from libraries absent from their training corpora. 
Our empirical study shows that simply exposing models to the required API knowledge is insufficient: even with oracle API specifications, LLMs still frequently fail to invoke private-library APIs correctly. 
To address this limitation, we proposed \name, which automatically synthesizes library-specific training data through Progressive Graph Evolution and Multidimensional Graph Pruning, enabling LLMs to learn how to use private libraries rather than merely see their documentation. 
To support rigorous evaluation, we constructed two benchmarks based on recently released libraries that are largely unfamiliar to modern LLMs. 
Experiments across multiple models show that \name substantially improves private-library-oriented code generation while largely preserving general code generation capability, and ablation studies further confirm the contribution of its key components.

\section{Data Availability}
Our code, benchmarks, and supplementary materials can be accessed through a public repository at: \url{https://github.com/THU-Agent/PriCoder}.

\section*{Acknowledgment}

This work was supported by the National Natural Science Foundation of China (No. 62602364), the Beijing Natural Science Foundation (No. 4264107), and the Tsinghua University–Keystone Electrical (Zhejiang) Co., Ltd. Joint Research Center for Embodied Multimodal Artificial Intelligence (JCEMAI).

\bibliographystyle{preprint}
\bibliography{ref}

\end{document}